\documentclass[reprint]{JASA}

\usepackage{placeins}
\usepackage[utf8]{inputenc}
\usepackage[english]{babel}
\usepackage{graphics}
\usepackage{graphicx}
\usepackage{amsmath,amssymb,esint}
\usepackage{multirow}
\usepackage{amsmath}
\usepackage{comment}
\usepackage{upgreek}
\usepackage{booktabs}
\usepackage{lineno}

\allowdisplaybreaks

\newcommand{\revone}[1]{{\color{black}{#1}}}
\newcommand{\revtwo}[1]{{\color{black}{#1}}}
\newcommand{\revall}[1]{{\color{black}{#1}}}

\begin{document}

\title[Acoustic wave modulation in accelerating flows]{Amplitude modulation of acoustic waves in accelerating flows quantified using acoustic black and white hole analogues}
\author{S\"{o}ren Schenke}
\affiliation{Chair of Mechanical Process Engineering, Otto-von-Guericke-Universit\"{a}t Magdeburg, Universit\"atsplatz 2, 39106 Magdeburg, Germany}

\author{Fabian Sewerin}
\affiliation{Chair of Mechanical Process Engineering, Otto-von-Guericke-Universit\"{a}t Magdeburg, Universit\"atsplatz 2, 39106 Magdeburg, Germany}

\author{Berend van Wachem}
\affiliation{Chair of Mechanical Process Engineering, Otto-von-Guericke-Universit\"{a}t Magdeburg, Universit\"atsplatz 2, 39106 Magdeburg, Germany}

\author{Fabian Denner}
\email{fabian.denner@polymtl.ca}
\affiliation{Department of Mechanical Engineering, Polytechnique Montr\'eal, Montr\'eal, H3T 1J4, QC, Canada}

\date{\today}

\begin{abstract}
We investigate the amplitude modulation of acoustic waves in accelerating flows, a problem that is still not fully understood, but essential to many technical applications, ranging from medical imaging to acoustic remote sensing. The proposed modeling framework is based on a convective form of the Kuznetsov equation, which incorporates the background flow field and is solved numerically by a finite-difference method. Using acoustic black and white hole analogues as model systems, we identify a modulation of the wave amplitude, that is shown to be driven by the divergence/convergence of the acoustic wave characteristics in an accelerating/decelerating flow, and which is distinct from the convective amplification accompanying an acoustic emitter moving at a constant velocity. To rationalize the observed amplitude modulation, a leading-order model is \revall{derived from first principles, leveraging} a similarity of the wave characteristics and the wave amplitude with respect to a modified Helmholtz number. This leading-order model may serve as a basis for the numerical prediction and analysis of the behavior of acoustic waves in accelerating flows, by taking advantage of the notion that any accelerating flow field can be described locally as a virtual acoustic black or white hole.
\end{abstract}


\maketitle


\section{Introduction}

The knowledge of how acoustic waves are modulated by the motion of wave emitters, wave-scattering boundaries or the wave carrier medium can be harnessed in many technical applications, ranging from acoustic remote sensing to medical Doppler ultrasound and imaging techniques. The waveform of the received acoustic signal encodes distinct information about a moving wave emitter/scatterer or the environment in which the wave has propagated. Prominent examples are Doppler ultrasound measurements of cardiac tissue motion and blood flows \citep{Oglat_et_al_2018} or the remote sensing of ocean currents \citep{Dowling_et_al_2015}. Similar concepts apply in the context of Doppler-modulated light and electromagnetic waves, for instance to investigate extraterrestrial atmospheric flows \citep{Showman_et_al_2013} or in the radar sensing of vital signs \citep{Li_et_al_2009}. 

While the classical Doppler shift of acoustic waves at uniform emitter or flow motion is predicted easily, the modulated waveform is considerably more difficult to interpret if either the wave emitter/scatterer \citep{Christov_2017} and/or the background medium \citep{Ewert_and_Schroeder_2003} are in non-uniform motion.
While significant progress has been made in understanding the Doppler-related generation of frequency side-bands \citep{Gasperini_et_al_2021}, the mechanisms leading to Doppler modulations of the acoustic wave amplitude appear to be more delicate \citep{Christov_2017}. Understanding the Doppler modulations becomes even more complicated if constitutive nonlinearities are involved, as for instance in medical high-intensity focused ultrasound applications \citep{Tabak_et_al_2022}. Due to the complexity of the problem, the interpretation of Doppler-modulated wave spectra increasingly relies on deep learning techniques \citep{Brooks_et_al_2018}, for which, however, advanced simulation tools are required to provide an adequate database \citep{Gasperini_et_al_2022}.

\begin{figure*}
\centering
\includegraphics[width=0.8\linewidth]{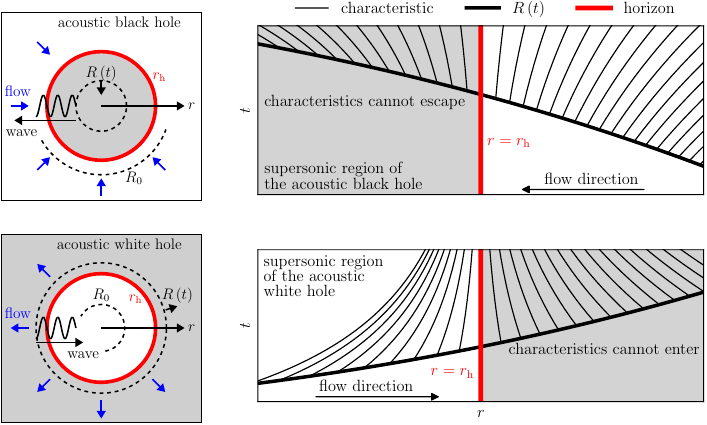}
\caption{Schematic of \revall{the spherically symmetric flow configurations and the characteristics of the considered} acoustic black hole (top) and acoustic white hole (bottom) analogues, where the moving wave-emitting boundary with instantaneous position $R(t)$ displaces the surrounding fluid such that a spatially non-uniform flow with a stationary sonic horizon $r_{\mathrm{h}}$, shown in red, is established. \revall{The figures on the left illustrate the spherically symmetric geometry with radial coordinate $r$ and the corresponding directions in which the wave-emitting boundary moves along with the flow, with the acoustic waves propagating opposite to the flow. $R_0$ denotes the initial position of the wave-emitting boundary. The figures on the right illustrate the evolution of the wave characteristics in the space-time ($r$-$t$) plane.}}
\label{fig:horizonSketch}
\end{figure*}

In this work, using a convective Kuznetsov equation that incorporates the background flow field and a moving wave-emitting boundary, solved numerically by a finite-difference method, we study the amplitude modulation of acoustic waves in accelerating flows. To this end, we use \revall{spherically symmetric} acoustic black hole \citep{Unruh_1981, Visser_1998} and acoustic white hole \citep{Mayoral_et_al_2011} analogues as model systems, \revall{as illustrated in Fig.~\ref{fig:horizonSketch}}, whereby we exploit the distinct geometrical features of the acoustic black/white hole \citep{Schenke_et_al_2022_PoF} and take advantage of the idea that any accelerating flow field can be transformed locally into a virtual acoustic black/white hole by a Galilean transformation. Both these model systems include a sonic transition at a fixed point in space, termed the sonic horizon, which allows us to investigate the behavior of a stationary acoustic wave in an accelerating flow field.

In the acoustic black hole configuration, outward propagating waves cannot escape to infinity from the supersonic flow region, whereas in the acoustic white hole configuration, inward propagating waves, starting in the subsonic flow region, cannot enter the supersonic region.
The kinematic features of acoustic wave propagation in an acoustic black or white hole configuration are described accurately in terms of a modified Helmholtz number, defined as \citep{Schenke_et_al_2022_PoF}
\begin{equation}
\mathrm{He} 
= \dfrac{c_0^2}{|a_{\mathrm{h}}| \lambda_{\mathrm{a}}} 
= \dfrac{c_0 f_{\mathrm{a}}}{|a_{\mathrm{h}}|},
\label{eq:HelmholtzABH}
\end{equation}
where $c_0$ is the speed of sound, $\lambda_\mathrm{a}$ and $f_\mathrm{a}$ are the length and frequency of the emitted acoustic waves, respectively, and $a_\mathrm{h}$ is the flow acceleration at the sonic horizon.
While our previous work demonstrated a geometric similarity of the emitted acoustic waves described by this Helmholtz number in accelerating flows, here we identify an additional dynamic similarity of the modulation of the wave amplitude in relation to Eq.~\eqref{eq:HelmholtzABH}. We leverage these similarities to build a leading-order model for the length and amplitude of acoustic waves in accelerating flows \revall{that can be used to assess whether acceleration-induced amplitude modulations play an important role in any application.}

In the following, Sec.~\ref{sec:Preliminary considerations} discusses preliminary considerations that further motivate the use of acoustic black and white hole analogues in this study. Subsequently, Sec.~\ref{sec:Mathematical model} introduces the employed modeling framework, \revall{Sec.~\ref{sec:Setup} discusses the considered model systems,} and Sec.~\ref{sec:Results} presents the results of representative acoustic black and white hole configurations. A leading-order model that incorporates the physical mechanisms underlying the modulation of the length and amplitude of acoustic waves in accelerating flows is presented in Sec.~\ref{sec:Local analytical model}, and the article is concluded in Sec.~\ref{sec:Conclusion}.

\section{Preliminary considerations}\label{sec:Preliminary considerations}

The following preliminary considerations are concerned with the geometrical aspects of the \revall{employed spherically symmetric} acoustic black and white hole analogues, discussed for a space-time plane with the spatial and temporal coordinates $r$ and $t$, as illustrated in Fig.~\ref{fig:horizonSketch}. \revall{The wave-emitting boundary of a spherical emitter,} with instantaneous position $R(t)$ \revall{of this boundary,} is thought to emit a continuous and initially monochromatic wave train. The wave characteristics evolve along the path
\begin{equation}
r_{\mathrm{c}}\left(t\right) = r_{\mathrm{c}}\left(t=0\right)  +  \int_0^t \left[ c_0(r(t),t) + u_0\left(r_{\mathrm{c}}(t),t\right) \right] \mathrm{d}t,
\label{eq:characteristic}
\end{equation}
where \revtwo{$c_0$ is the speed of sound of the background medium}, $u_0$ is the background flow velocity and where the integral is evaluated numerically to obtain the characteristics in Fig.~\ref{fig:horizonSketch}. The wave-emitting boundary displaces the background medium and collapses (acoustic black hole) or expands (acoustic white hole) in such a way that a stationary, \revall{spherically symmetric} background flow with a sonic transition is obtained, \revall{which} can be associated with a potential sink or source of constant strength, \revall{as further detailed in Sec.~\ref{sec:Setup}}. The point of sonic transition at $r=r_{\mathrm{h}}$, highlighted in red in Fig.~\ref{fig:horizonSketch}, is referred to as the \textit{sonic horizon} and has the particular feature that the outgoing characteristic in the acoustic black hole configuration and the ingoing characteristic in the acoustic white hole configuration are stationary at this point.
Both the background flow velocity and acceleration are constant \revall{at the sonic horizon and, in turn, along the stationary characteristic situated at the sonic horizon, which} allows to isolate the effect of the background flow acceleration on the wave amplitude at the stationary sonic horizon.
These acoustic black and white hole analogues, hence, greatly simplify the study of acoustic waves in an accelerating flow.

While the presence of the sonic horizon formally requires a sonic transition, the concept of the {\it virtual acoustic black hole} is now introduced. In this concept, an accelerating flow field of \revone{arbitrary} (also subsonic) velocity is observed in a stationary frame of reference. It is further assumed that a reference characteristic exists, for which the background flow velocity and the background flow acceleration take {\it constant} values along its path in the space-time plane for an arbitrarily small time increment, so that the curvature of the reference characteristic vanishes, whereas the neighboring characteristics are deflected by an accelerating background flow along their paths. The idea is that \revone{a} characteristic of vanishing curvature can be associated with a stationary apparent sonic horizon, by observing it from a moving inertial frame of reference in which the magnitude of the constant relative flow velocity along the characteristic matches the speed of sound, thus forming an apparent sonic horizon under the appropriate Galilean transformation. 
Since the conservation equations for mass and momentum are invariant under Galilean transformation in the non-relativistic limit \citep{Wang_2022}, the acoustic pressure modulation along a given characteristic is not affected by the Galilean transformation.
Consequently, the virtual acoustic black hole, representing diverging wave characteristics, together with its counterpart, the virtual acoustic white hole, representing converging wave characteristics, may represent any accelerating flow.

These preliminary considerations illustrate the role of the acoustic black and white hole analogues as model systems for accelerating flows, with the beneficial geometrical feature that one particular characteristic is trapped at the fixed observation point $r_{\mathrm{h}}$, \revtwo{i.e.,~at the sonic horizon,} subject to a constant background flow velocity and acceleration.

\section{Modeling framework}\label{sec:Mathematical model}

The employed modeling framework is based upon a quasi one-dimensional second-order acoustic wave equation for transient and spatially non-uniform background flows, derived from first principles by a perturbation of the conservation equations governing the fluid flow \citep{Unruh_1981, Visser_1998, Ewert_and_Schroeder_2003}. Consistently retaining second-order acoustic terms, and introducing second-order expansions of the linearized barotropic equation of state \citep{Shevchenko_and_Kaltenbacher_2015} and the linearized fluid enthalpy \citep{Prosperetti_1984}, we arrive at what can be seen as a lossless and convective form of the Kuznetsov equation \citep{Kuznetsov_1971}. Among the second-order acoustic wave equations, the Kuznetsov equation is often considered as the physically most accurate one \citep{Dekkers_et_al_2020}, and its formal range of validity includes curved wavefronts \citep{Gu_and_Jing_2015}, as considered in this work.

\subsection{Modeling assumptions and conventions}\label{sec:Modeling assumptions and conventions}

The governing equations are formulated for an inviscid quasi one-dimensional and, hence, irrotational flow with spatial coordinate $r$. The flow is assumed to pass through a spatially variable cross-sectional area $A\left(r\right)$, with the Laplace operator given by
\begin{equation}
\nabla_{\mathrm{A}}^2 =
\dfrac{\partial^2}{\partial r^2} + \dfrac{1}{A}\dfrac{\partial A}{\partial
r}\dfrac{\partial}{\partial r}.
\label{eq:LaplacianDef}
\end{equation}
The continuity equation for this flow reads 
\begin{equation}
\dfrac{\partial \rho}{\partial t} + \dfrac{1}{A}\dfrac{\partial}{\partial r}\left(A\rho u\right) = 0,
\label{eq:Conti}
\end{equation}
where $\rho$ and $u$ are the density and velocity of the fluid, and where $u$ follows from the velocity potential $\phi$ as $u = \partial \phi/\partial r$. The velocity potential $\phi$, the velocity $u$, the pressure $p$, and the density $\rho$ are formally decomposed into background contributions and acoustic perturbations by writing them in terms of the first-order series $\phi = \phi_0 + \epsilon\phi_1$, $u = u_0 + \epsilon u_1$, $p = p_0 + \epsilon p_1$, and $\rho = \rho_0 + \epsilon\rho_1$, where the subscript $0$ indicates the background state, which is assumed to be known, and the subscript $1$ indicates the unknown acoustic perturbation \citep{Visser_1998}. The dimensionless quantity $\epsilon$ is constant and equal to unity, and it is formally introduced to indicate the order of nonlinearity of the acoustic terms in the course of the following derivations.
\revtwo{It is assumed that the fluid is barotropic, with $\mathrm{d} p/\mathrm{d} \rho = c_0^2 \approx p_1/\rho_1$, and that all compressibility effects are captured by the acoustic perturbations. Consequently, the background flow is assumed to be incompressible, with a constant density $\rho_0$ and speed of sound $c_0$. The background flow field $u_0\left(r,t\right)$, however, may be both spatially non-uniform and time-dependent. The material derivative operator is defined based on the background flow velocity,}
\begin{equation}
\dfrac{\mathrm{D}}{\mathrm{D}t} = \dfrac{\partial}{\partial t} + u_0 \dfrac{\partial}{\partial r},
\label{eq:MaterialDerivative}
\end{equation}
and the Lagrangian density is given by \citep{Cervenka_and_Bednarik_2022}
\begin{equation}
\mathcal{L} = \dfrac{1}{2}\rho_0u_1^2 - \dfrac{1}{2}\dfrac{p_1^2}{\rho_0 c_0^2}
\label{eq:LagrangianDensity}.
\end{equation}

As discussed by \citet{Cervenka_and_Bednarik_2022}, the unknown nonlinear relation between $p_1$ and $u_1$ requires to formulate the second-order acoustic wave equation, i.e.,~the Kutznetsov equation, in terms of the acoustic potential $\phi_1$, where the relation $u_1 = \partial\phi_1/\partial r$ can be employed. The acoustic pressure $p_1$ is then obtained by consulting the conservation of energy or momentum.

\subsection{Lossless convective Kuznetsov equation}\label{sec:Lossless Convective Kuznetsov equation}

Following the suggestion of \citet{Visser_1998}, the potential form of the transient Bernoulli equation \citep{Prosperetti_1984},
\begin{equation}
\dfrac{\partial \phi}{\partial t} + \dfrac{1}{2}\left(\dfrac{\partial \phi}{\partial r}\right)^2 + h\left(p\right) = 0,
\label{eq:Bernoulli}
\end{equation}
where $h\left(p\right) = \int_{p_0}^p \rho^{-1} \text{d}p$ is the enthalpy difference \revtwo{related to the local pressure $p$ and the background pressure $p_0$}, is used as the starting point for the derivation of the lossless convective Kuznetsov equation. 
Substituting the perturbations introduced in Sec.~\ref{sec:Modeling assumptions and conventions} into Eqs.~\eqref{eq:Conti} and \eqref{eq:Bernoulli}, and isolating the acoustic terms of order $\mathcal{O}(\epsilon^1)$ and $\mathcal{O}(\epsilon^2)$, the conservation equations for the acoustic perturbations are obtained as
\begin{eqnarray}
\dfrac{\partial \rho_1}{\partial t} + \dfrac{1}{A}\dfrac{\partial}{\partial r}\left[A\left(\rho_0 u_1 + \rho_1 u_0 + \rho_1 u_1\right)\right] &=& 0,
\label{eq:Conti1} \\
\revall{\dfrac{\mathrm{D} \phi_1}{\mathrm{D} t} + \dfrac{1}{2}\left(\dfrac{\partial \phi_1}{\partial r}\right)^2 + h_1} &=& 0,
\label{eq:Bernoulli_interim}
\end{eqnarray}
\revall{where $\partial \phi_0/\partial r = u_0$ by definition and the acoustic contribution to the enthalpy difference is given as \citep{Prosperetti_1984} 
\begin{equation}
h_1 = \dfrac{p_1}{\rho_0} - \dfrac{1}{2c_0^2}\left(\dfrac{p_1}{\rho_0}\right)^2.
\label{eq:hexpansion}    
\end{equation}
The constitutive nonlinearity of the medium is taken into account by expanding the acoustic pressure state to second order, such that \citep{Shevchenko_and_Kaltenbacher_2015}
\begin{equation}
p_1 = \rho_1 c_0^2 - \dfrac{B}{2A}\dfrac{\rho_1^2 c_0^2}{\rho_0},
\label{eq:pexpansion}
\end{equation}
where $B/(2A)$ is the nonlinearity parameter. The mutual interaction between the acoustic pressure and the acoustic (particle) velocity, required to convert between the acoustic velocity
potential $\phi_1$ and the acoustic pressure $p_1$, is then given as 
\begin{equation}
\dfrac{p_1}{\rho_0} = - \dfrac{\mathrm{D}\phi_1}{\mathrm{D}t} - \dfrac{\mathcal{L}}{\rho_0} - \dfrac{\beta - 1}{c_0^2}\left(\dfrac{\mathrm{D}\phi_1}{\mathrm{D}t}\right)^2,
\label{eq:rhoExpansion}
\end{equation}
where $\beta = 1 + B/(2A)$ is the nonlinearity coefficient.
With the barotropic assumption, $\rho_1 = p_1/c_0^2$, and retaining only terms up to second order yields, after some lengthy rearrangement, the lossless convective Kuznetsov equation
\begin{align}
&\left(1 + \dfrac{2\beta - 1}{c_0^2}\dfrac{\mathrm{D}\phi_1}{\mathrm{D}t}\right)\dfrac{\mathrm{D}^2\phi_1}{\mathrm{D}t^2} 
- c_0^2\nabla_{\mathrm{A}}^2\phi_1 \nonumber \\
&+ \dfrac{1}{\rho_0}\dfrac{\mathrm{D}\mathcal{L}}{\mathrm{D}t}
+ \dfrac{\partial \phi_1}{\partial r}\dfrac{\partial}{\partial r}\left(\dfrac{\mathrm{D}\phi_1}{\mathrm{D}t}\right) = 0,
\label{eq:convectiveKuznetsov}
\end{align}
where the potential form of the Lagrangian density $\mathcal{L}$ is defined as
\begin{equation}
\mathcal{L} = \dfrac{\rho_0}{2}\left(\dfrac{\partial\phi_1}{\partial r}\right)^2 - \dfrac{\rho_0}{2c_0^2}\left(\dfrac{\mathrm{D}\phi_1}{\mathrm{D} t}\right)^2.
\label{eq:LagrangianDensity2}
\end{equation}
A detailed step-by-step derivation of the lossless convective Kuznetsov equation is provided in the supplementary material.}

The lossless convective Kuznetsov equation proposed in Eq.~\eqref{eq:convectiveKuznetsov} does not make any assumptions about the origin of the background flow field, meaning that any flow field, also resulting from experimental measurements or numerical simulations, may be considered. Eq.~\eqref{eq:convectiveKuznetsov} reduces to the quasi one-dimensional and lossless form of the standard Kuznetsov equation \citep{Kuznetsov_1971, Cervenka_and_Bednarik_2019} for $u_0 = 0$, with $\mathrm{D/\mathrm{D}}t = \partial/\partial t$. If the second-order nonlinear terms are neglected and assuming a non-zero background flow field, Eq.~\eqref{eq:convectiveKuznetsov} reduces to the convective wave equation, $\mathrm{D}^2\phi_1/\mathrm{D}t^2 - c_0^2\nabla_{\mathrm{A}}^2\phi_1 = 0$. 

\subsection{Coordinate transformation}\label{sec:Moving wave-emitting domain boundary}

\revall{The mathematical modeling and numerical description of the moving wave-emitting domain boundary and the ensuing background flow pose two specific difficulties. First, the governing wave equation, Eq.~\eqref{eq:convectiveKuznetsov}, is ill-posed at the sonic transition.
This becomes readily apparent by separating the first term on the left-hand side of Eq.~\eqref{eq:convectiveKuznetsov} into a linear and a nonlinear term with respect to $\phi_1$.
Expanding the resulting linear term, $\mathrm{D}^2\phi_1/\mathrm{D}t^2$, using Eq.~\eqref{eq:MaterialDerivative}, and expanding the Laplace operator using Eq.~\eqref{eq:LaplacianDef}, Eq.~\eqref{eq:convectiveKuznetsov} can be rewritten as
\begin{eqnarray}
\dfrac{\partial^2\phi_1}{\partial t^2} 
- \left(c_0^2-u_0^2\right)\dfrac{\partial^2\phi_1}{\partial r^2}
+ 2u_0\dfrac{\partial^2\phi_1}{\partial r\partial t}
+ \left(\dfrac{\mathrm{D}u_0}{\mathrm{D}t} - \dfrac{c_0^2}{A}\dfrac{\partial A}{\partial r}\right)\dfrac{\partial \phi_1}{\partial r}
\nonumber \\
=
- \dfrac{\partial}{\partial r}\left(\dfrac{\mathrm{D}\phi_1}{\mathrm{D}t}\right)\dfrac{\partial \phi_1}{\partial r}
- \dfrac{1}{\rho_0}\dfrac{\mathrm{D}\mathcal{L}}{\mathrm{D}t}
- \dfrac{2\beta - 1}{c_0^2}\dfrac{\mathrm{D}\phi_1}{\mathrm{D}t}\dfrac{\mathrm{D}^2\phi_1}{\mathrm{D}t^2}.
\label{eq:convectiveKuznetsov_partExpanded}
\end{eqnarray}
The Laplacian coefficient $(c_0^2-u_0^2)$ on the left-hand side causes a sign change at the point of sonic transition that renders the lossless convective Kuznetsov equation ill-posed, similar to the first-order convective wave equation used in our previous work \citep{Schenke_et_al_2022_PoF}. Second, the wave-emitting boundary moves relative to the desired stationary observation point represented by the stationary sonic horizon.
To resolve both difficulties, a coordinate transformation between the time-varying physical domain and a fixed computational domain is employed.}

\begin{figure}
{\includegraphics[width=0.95\reprintcolumnwidth]{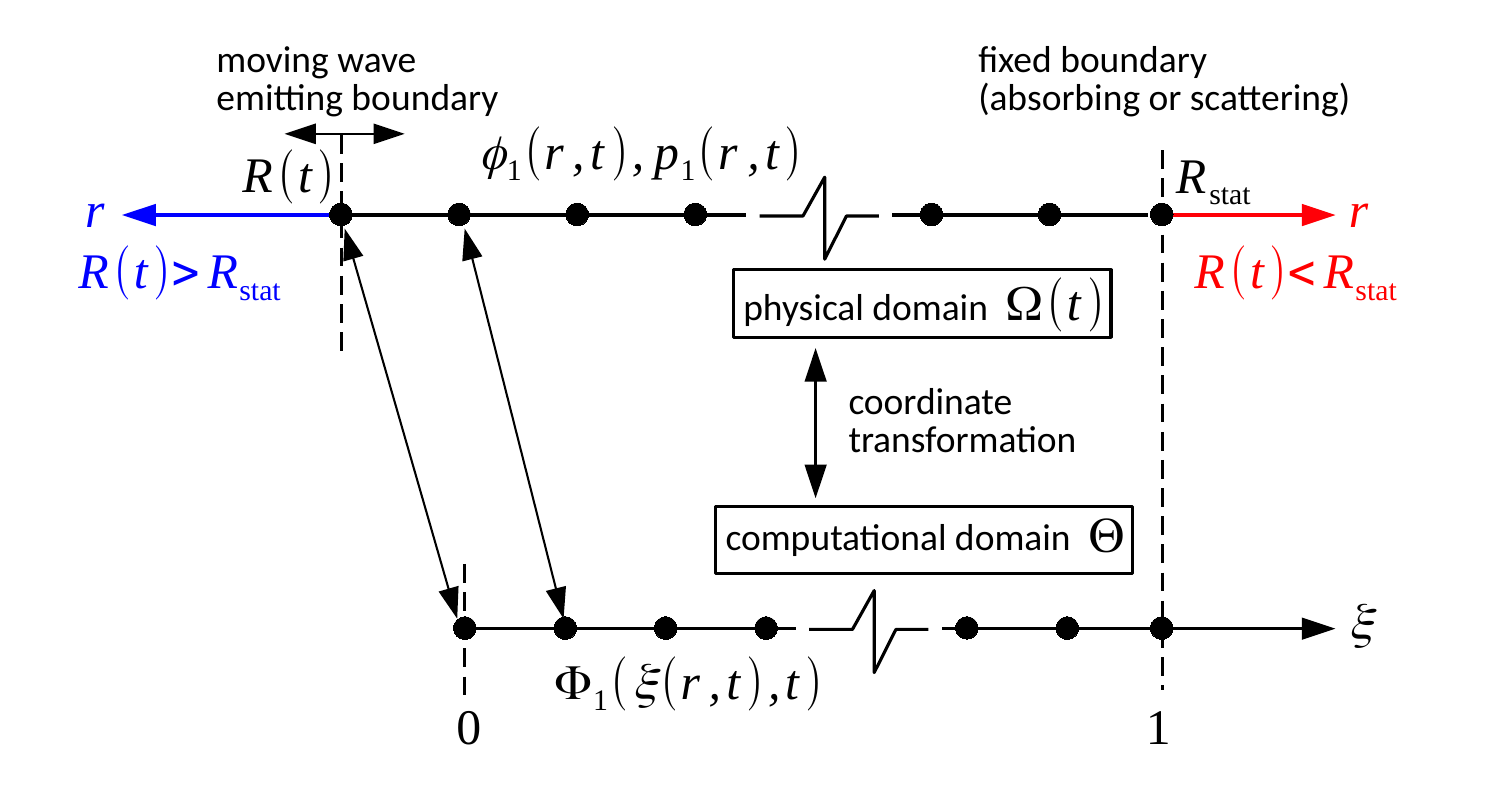}}
\caption{Schematic of the physical domain $\Omega\left(t\right)$, with moving wave-emitting boundary $R\left(t\right)$ and fixed boundary $R_{\mathrm{stat}}$, and the fixed computational domain $\Theta$. The $r$-axis is defined as positive towards the right if $R\left(t\right)<R_{\mathrm{stat}}$ (shown in red) and as positive towards the left if $R\left(t\right)>R_{\mathrm{stat}}$ (shown in blue), so that the moving boundary $r=R\left(t\right)$ in $\Omega\left(t\right)$ always maps to $\xi=0$ in $\Theta$.}
\label{fig:domain3}
\end{figure}

The position of the wave-emitting domain boundary is assumed to be time-dependent. Similar to the approaches by \citet{Christov_2017} and \citet{Gasperini_et_al_2021}, we invoke a time-dependent coordinate transformation \citep{Schenke_et_al_2022}
\begin{equation}
r: \: \left[\mathcal{X}_R,\mathcal{X}_{\infty}\right]
\times \left[0, \infty\right)\rightarrow \left[R\left(t\right),R_{\mathrm{stat}}\right], \: \left(\xi,t\right) \mapsto r\left(\xi,t\right)
\label{eq:coordTrans}
\end{equation}
between a moving physical domain $\Omega\left(t\right)$ with the time-dependent left boundary $R\left(t\right)$ and a fixed right boundary $R_{\mathrm{stat}}$, and a fixed computational domain $\Theta$ with \revtwo{spatial coordinate $\xi$ and} fixed left and right boundaries $\mathcal{X}_R$ and $\mathcal{X}_{\infty}$, in which the transformed governing equation is solved by means of standard finite-difference techniques. Without loss of generality, the fixed domain boundaries of $\Theta$ are chosen to be $\mathcal{X}_R=0$ and $\mathcal{X}_{\infty}=1$. As illustrated in Fig.~\ref{fig:domain3}, the coordinate transformation conveys a mapping between the two domains, such that $\phi_1\left(r\left(\xi,t\right),t\right)=\Phi_1\left(\xi,t\right)$. The $k$-th spatial and temporal derivatives \revtwo{of the acoustic potential $\phi_1$} can then be written as
\begin{eqnarray}
\dfrac{\partial^k \phi_1}{\partial r^k} &=& \left(\dfrac{\partial \xi}{\partial r} \dfrac{\partial}{\partial \xi}\right)^k\Phi_1,
\label{eq:transDerivativeOp_space} \\
\dfrac{\partial^k \phi_1}{\partial t^k} &=& \left(\dfrac{\partial \xi}{\partial t} \dfrac{\partial}{\partial \xi} + \dfrac{\partial}{\partial t}\right)^k\Phi_1.
\label{eq:transDerivativeOp_time}
\end{eqnarray}
The derivatives of $\xi$ depend on the employed coordinate transformation. We opt for the linear transformation
\begin{equation}
\xi\left(r,t\right) = \mathcal{X}_{\infty} + \left(r - R_{\mathrm{stat}}\right) \dfrac{\mathcal{X}_{\mathrm{\infty}} - \mathcal{X}_{R}}{R_{\mathrm{stat}}-R\left(t\right)}.
\label{eq:linTrans}
\end{equation}
The change of variables for Eqs.~\eqref{eq:rhoExpansion} and \eqref{eq:convectiveKuznetsov}, as well as the derivatives of $\xi$ that follow from Eq.~\eqref{eq:linTrans}, are \revall{detailed in the supplementary material.}

Eq.~\eqref{eq:linTrans} represents a Galilean-type transformation, i.e.,~a time-dependent transformation in physical space. However, it differs from the standard Galilean transformation used by \citet{Christov_2017} and \citet{Gasperini_et_al_2021} in that it conveys a dilation of the physical domain, as indicated in Fig.~\ref{fig:domain3}, rather than a pure translation. The dilation results in a linear velocity distribution for the grid points in $\Omega\left(t\right)$. This velocity distribution plays an important role in preserving the well-posedness of the transformed convective Kuznetsov equation, in the presence of a conceptually transonic flow field \citep{Schenke_et_al_2022_PoF}. 

For the problem to be well-posed in the considered frame of reference, it is essential that a near-invariant of the standard wave equation is obtained in the sense that the sign of the Laplacian term in Eq.~\eqref{eq:convectiveKuznetsov} is preserved \citep{Christov_2017}.
As pointed out by \citet{Gregory_et_al_2015}, an invariant of the standard wave equation is obtained by transforming the convective wave equation into the fluid's frame of reference. In this regard, the grid points in $\Omega\left(t\right)$ should ideally follow the background flow \citep{Schenke_et_al_2022_PoF}. Even though this cannot be strictly guaranteed for an accelerating flow field, due to the linearity of Eq.~\eqref{eq:linTrans}, the dilation of $\Omega\left(t\right)$ can, at least within certain limits, preserve the sign of the Laplacian coefficient of the transformed convective wave equation.

\subsection{Numerical method}

A spatially uniform grid is used to discretize the transformed governing equations.
The applied explicit finite-difference method \citep{Schenke_et_al_2022} is a standard procedure based on central differences in space, complemented by a predictor-corrector method adopted from the works of \citet{Dey_and_Dey_1983} and \citet{Nascimento_et_al_2010} to suppress the formation of dispersive numerical noise in the presence of shocks and frequency side-bands. The corrected solution is weighted by a factor $\gamma$ that controls the magnitude of the damping imposed by the corrector step \citep{Schenke_et_al_2022}.
As indicated in Fig.~\ref{fig:domain3}, a wave-absorbing or wave-scattering boundary condition may be applied at the fixed boundary of the domain. However, as the waves do not reach the fixed boundary in the presented test cases, the choice of the boundary condition is immaterial in the present work.

\section{Model systems}
\label{sec:Setup}

For the further course of this study, we consider acoustic black and white hole configurations with a spherically symmetric and stationary background flow field that exhibits a stationary sonic horizon, \revall{see Fig.~\ref{fig:horizonSketch}}. In accordance with the virtual acoustic black and white hole analogues discussed in Sec.~\ref{sec:Preliminary considerations}, the transonic flow field may be associated with any flow field in a different inertial frame of reference, which allows us to assume a smooth sonic transition. The origin of the flow field in this alternative frame of reference is immaterial to the discussions in the present work, since the current study focuses on the acoustic wave dynamics along curved characteristics in an accelerating background flow, irrespective of how the acceleration of the flow is achieved. 

\revone{The spherically symmetric velocity field induced by the motion of the wave-emitting boundary can be decomposed into three distinct contributions \citep{Denner2023}: (i) the incompressible displacement of the fluid by the moving boundary, (ii) the compression/expansion of the fluid resulting from its displacement, and (iii) the particle velocity associated with the emitted acoustic waves. Since we assume that the displacement of the fluid does not compress or expand the fluid (i.e.,~$\rho_0 = \mathrm{const.}$ and $c_0 = \mathrm{const.}$) and acoustic velocity contributions are incorporated in the governing wave equation, the background flow  is constituted only by the incompressible displacement of the fluid, the velocity of which follows from the continuity equation as $u_0(r,t) = \dot{R}(t) \, R(t)^2/r^2$, where $R$ and $\dot{R}$ denote the position and velocity of the wave-emitting boundary.} In order to establish a stationary background flow field, \revtwo{$R(t)$ and $\dot R(t)$ are defined such that $\dot R(t)R(t)^2 = \mathrm{const.} = \mp c_0r_{\mathrm{h}}^2$, where the negative (positive) sign holds for the acoustic black (white) hole configuration, and where $c_0$ denotes the constant speed of sound of the background flow.} The solution of this ordinary differential equation readily yields the kinematic parameters 
\citep{Schenke_et_al_2022_PoF} \revtwo{$R(t) = (R_0^{3} \mp 3 c_0 r_{\mathrm{h}}^2 t)^{\frac{1}{3}}$, $\dot R(t) = \mp c_0r_{\mathrm{h}}^2 \, (R_0^{3} \mp 3 c_0 r_{\mathrm{h}}^2 t)^{-\frac{2}{3}}$, and $\ddot R\left(t\right) = -2 c_0^2 r_{\mathrm{h}}^{4} \, (R_0^{3} \mp 3 c_0 r_{\mathrm{h}}^2 t)^{-\frac{5}{3}}$.}
\revtwo{As a consequence, a stationary background flow is established, with} 
\begin{equation}
    \revtwo{u_0(r) = \dot{R} \, \frac{R^2}{r^2} = \mp c_0 \, \frac{r_{\mathrm{h}}^2}{r^2}.} \label{eq:u0_spherical}
\end{equation}
\revtwo{At the sonic horizon $r_\mathrm{h}$, where $u_0 (r=r_{\mathrm{h}}) = \mp c_0$, the steady-state acceleration follows as}
\begin{equation}
 a_{\mathrm{h}} =  \mp \left( u_0 \dfrac{\partial u_0}{\partial r}\right)_{r_{\mathrm{h}}} = \pm \dfrac{2 c_0^2}{r_{\mathrm{h}}} = \mathrm{const.},
\label{eq:gh}
\end{equation}
being positive (negative) for the acoustic black (white) hole configuration. The acceleration may be defined more generally as $a_{\mathrm{h}} = \hat{\mathbf{u}} \cdot \left(\mathbf{u}_0 \cdot \nabla \mathbf{u}_0 \right)_\mathrm{r_\mathrm{h}}$, where $\hat{\mathbf{u}} = \mathbf{u}_0/|\mathbf{u}_0|$ is the unit vector of the background flow direction.

\revtwo{The sonic horizon of the of the acoustic black and white hole configurations is characterized by a background flow that varies in space but not in time. At the sonic horizon the flow does not change and, consequently, the density and speed of sound are constant, justifying the assumption of an incompressible background flow with shock-free sonic transition.
The sonic transition at the sonic horizon is only relevant to the parameterization of the emitted acoustic waves with respect to the background flow acceleration, employing the modified Helmholtz number given in Eq.~\eqref{eq:HelmholtzABH}, as well as the sonic horizon as a well-defined observation point for the wave amplitude modulation \citep{Schenke_et_al_2022_PoF}. The magnitude of the velocity field is irrelevant to the investigated acceleration-driven amplitude modulation of acoustic waves at a particular observation point, such as the sonic horizon, as discussed in Sec.~\ref{sec:Preliminary considerations}.
However, the validity of the presented results becomes increasingly inaccurate with increasing distance from the sonic horizon, since an expanding range of Mach numbers is traversed by an acoustic wave.}

The considered speed of sound and density of the background fluid are representative of water, with $c_0=1500\mathrm{\:m/s}$ and $\rho_0=1000\mathrm{\:kg/m^3}$, respectively. 
The excitation pressure amplitude $\Delta p_{\mathrm{a}}$ is chosen such that the nominal shock formation distance of the unmodulated nonlinear wave in a quiescent background medium and in one-dimensional Cartesian coordinates \citep{Blackstock_1966} is $\Delta r_{\mathrm{sh}}=35\lambda_{\mathrm{a}}$. The number of grid points per emitted wavelength is $900$, the time-step $\Delta t$ satisfies $c_0 \Delta t / \Delta r = 0.225$ with respect to the initially uniform mesh spacing $\Delta r$, and the weight of the corrector step in the predictor-corrector method is $\gamma=1$.

\section{Results}
\label{sec:Results}

\revall{Representative results for the considered acoustic black and white hole configurations are presented below. The modeling framework proposed in Section \ref{sec:Mathematical model} and the model systems described in Section \ref{sec:Setup} are} implemented in the open-source software program {\tt Wave-DNA}, of which the version (v1.1) used to produce the presented results is available at \url{https://doi.org/10.5281/zenodo.8084168}, \revall{allowing a straightforward reproduction of the presented results.}

\subsection{Acoustic black hole configuration}\label{sec:Acoustic black hole configuration}

\begin{figure*}
    \includegraphics[width=\linewidth]{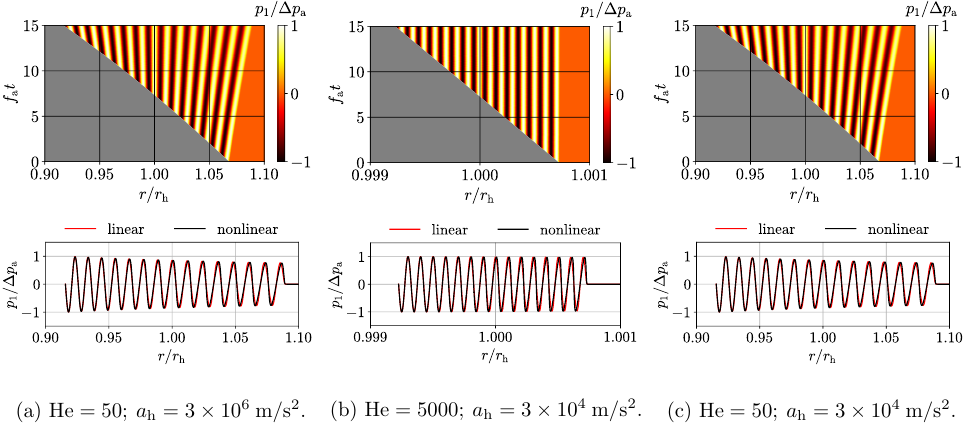}
    \caption{Space-time diagrams (top) and instantaneous wave profiles at $f_\mathrm{a} t=15$ (bottom) of the acoustic pressure $p_1$ in \revall{a spherically symmetric}  acoustic black hole configuration for a systematic variation of the Helmholtz number $\mathrm{He}$, given by Eq.~\eqref{eq:HelmholtzABH}, and the acceleration $a_{\mathrm{h}}$, given by Eq.~\eqref{eq:gh}.}
    \label{fig:ABH}
\end{figure*}

Fig.~\ref{fig:ABH} shows the acoustic pressure distribution in space-time diagrams alongside the instantaneous wave profiles for three acoustic black holes. 
\revall{Initially, the entire physical domain and, consequently, the moving wave-emitting boundary are located outside the acoustic black hole, see Fig.~\ref{fig:horizonSketch}.}
Subsequently, the wave-emitting boundary collapses and, eventually, passes the sonic horizon. The initial position $R\left(t=0\right)$ is calibrated such that out of 15 emitted wave periods, the positive peak of the eighth period coincides with the sonic horizon.

\begin{figure}
\includegraphics[width=0.95\reprintcolumnwidth]{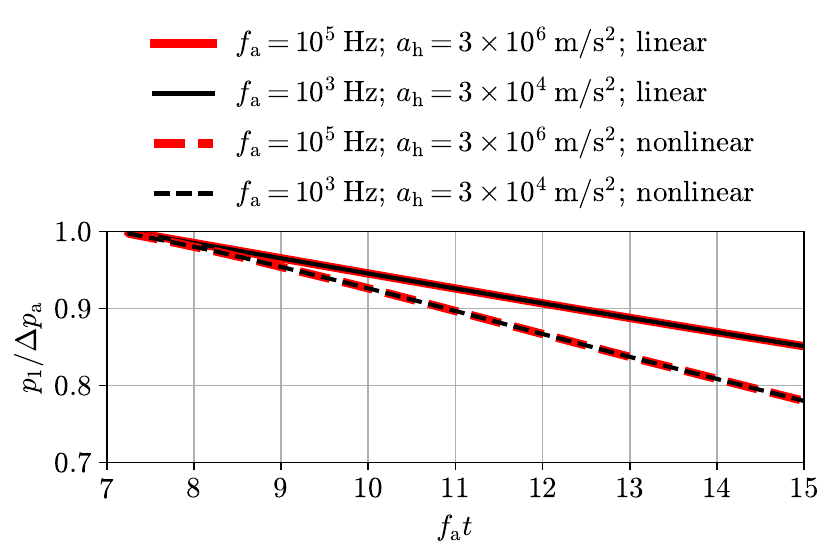}
\caption{Temporal evolution of the acoustic pressure $p_1$ at the sonic horizon $r_{\mathrm{h}}$ for the acoustic black holes with $\mathrm{He}=50$ in Figs.~\ref{fig:ABH}(a) and \ref{fig:ABH}(c).}
\label{fig:horizon_simABH}
\end{figure}

Starting from the acoustic black hole in Fig.~\ref{fig:ABH}(a) with sonic horizon radius $r_{\mathrm{h}}=\mathrm{1.5\:m}$, emitted wave frequency $f_{\mathrm{a}}=100\mathrm{\:kHz}$, and Helmholtz number $\mathrm{He}=50$, Fig.~\ref{fig:ABH}(b) is obtained by increasing the sonic horizon radius by a factor of 100 to $r_{\mathrm{h}}=\mathrm{150.0\:m}$ while keeping $f_{\mathrm{a}}$ and the physical simulation time constant, resulting in $\mathrm{He}=5000$. The flow acceleration as defined by Eq.~\eqref{eq:gh} implies a larger acceleration $a_{\mathrm{h}}$ for smaller horizon radii $r_{\mathrm{h}}$ and, consequently, a more rapidly varying background flow velocity. This causes the characteristics in Fig.~\ref{fig:ABH}(a) to diverge more rapidly than in Fig.~\ref{fig:ABH}(b). The configuration shown in Fig.~\ref{fig:ABH}(c) is obtained by turning the acoustic black hole shown in Fig.~\ref{fig:ABH}(b) into an acoustic black hole that is geometrically similar to the one in Fig.~\ref{fig:ABH}(a), but which has the same sonic horizon radius as in Fig.~\ref{fig:ABH}(b). This is achieved by decreasing the emitted wave frequency (increasing the emitted wavelength) by the ratio of the sonic horizon radii, i.e.,~$1.5/150.0$, so that $f_{\mathrm{a}}=1\mathrm{\:kHz}$ and $\mathrm{He}=50$ in Fig.~\ref{fig:ABH}(c). In order to simulate the same number of emitted wave periods, the physical simulation time is increased by the inverse of this ratio, hence a factor of 100. The rationale behind this geometric similarity observed between Figs. \ref{fig:ABH}(a) and \ref{fig:ABH}(c) is that in the less rapidly varying background flow field, the wave characteristics are given more space and time to sweep the same difference of $u_0$ as within a shorter duration in a more rapidly varying flow field \citep{Schenke_et_al_2022_PoF}.

\begin{figure*}
    \includegraphics[width=\linewidth]{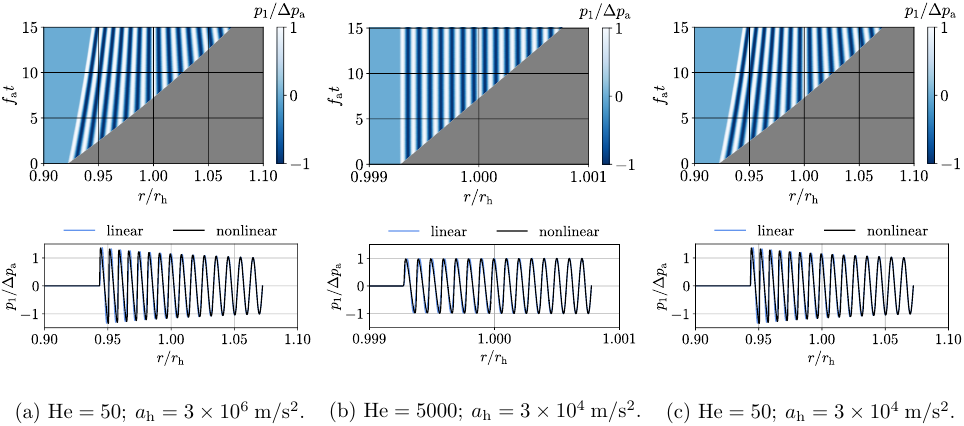}
    \caption{Space-time diagrams (top) and instantaneous wave profiles at $f_\mathrm{a} t=15$ (bottom) of the acoustic pressure $p_1$ in \revall{a spherically symmetric} acoustic white hole configuration for a systematic variation of the Helmholtz number $\mathrm{He}$, given by Eq.~\eqref{eq:HelmholtzABH}, and the acceleration $a_{\mathrm{h}}$, given by Eq.~\eqref{eq:gh}. }
    \label{fig:AWH}
\end{figure*}

While the space-time diagrams in Fig.~\ref{fig:ABH} are obtained for linear waves ($\beta=0$, $\mathcal{L}=0$), the bottom part of Fig.~\ref{fig:ABH} shows the instantaneous wave profiles of $p_1$ recorded at $f_{\mathrm{a}}t=15$, for both linear and nonlinear waves ($\beta=3.5$). The nonlinear waves exhibit a progressive steepening due to the constitutive nonlinearity of the fluid. For $\mathrm{He}=5000$ in Fig.~\ref{fig:ABH}(b), the wave amplitude remains virtually constant over $r$. For $\mathrm{He}=50$ in Figs.~\ref{fig:ABH}(a) and \ref{fig:ABH}(c), both the linear and the nonlinear wave decay in the radial direction. A closer inspection of Figs.~\ref{fig:ABH}(a) and \ref{fig:ABH}(c) shows that the wave amplitude decays along the outgoing wave characteristics, accompanied by a progressive divergence of the wave characteristics.
To further quantify the decay of the emitted waves, Fig.~\ref{fig:horizon_simABH} shows the evolution of the acoustic pressure $p_1$ at the sonic horizon position $r_{\mathrm{h}}$ for the geometrically similar acoustic black holes in Figs.~\ref{fig:ABH}(a) and \ref{fig:ABH}(c), for both the linear and nonlinear waves. Cases with identical Helmholtz numbers lead to the same decay of the acoustic pressure $p_1$ over the dimensionless time $f_{\mathrm{a}}t$, keeping in mind that the compared waves are emitted with different frequencies $f_{\mathrm{a}}$.  

Fig.~\ref{fig:horizon_simABH} \revall{seems to suggest} that the acoustic pressure $p_1$ decays faster for the nonlinear wave than for the linear wave. \revall{However,} in this particular case,  this \revall{apparent difference in amplitude} is not the result of the wave decay, but due to the nonlinear distortion of the wave profile, which results from a variation of the local speed of sound about $c_0$. As per the applied modeling assumptions, the magnitude of this variation is proportional to the acoustic pressure $p_1$ \citep{Blackstock_1966, Shevchenko_and_Kaltenbacher_2015}. \revall{As a consequence of this wave distortion, the peak of the nonlinear wave is tilted away from the sonic horizon and the pressure amplitude recorded at the sonic horizon ($r=r_\mathrm{h}$) reduces.}  Because this relation is linear, the progressive distortion of the wave profile is geometrically similar for flow-wave configurations of identical Helmholtz numbers. Presumably, this similarity breaks down if higher-order ($>2$) constitutive nonlinearities are taken into account, where the wave distortion cannot be expected to depend linearly on time. 

\subsection{Acoustic white hole configuration}\label{sec:Acoustic white hole configuration}

\begin{figure}
\includegraphics[width=0.95\reprintcolumnwidth]{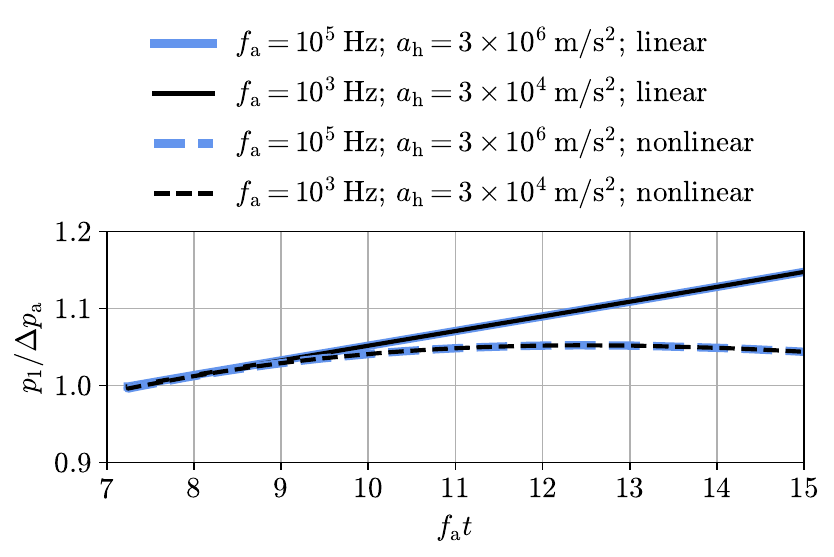}
\caption{Temporal evolution of the acoustic pressure $p_1$ at the sonic horizon $r_{\mathrm{h}}$ for the acoustic white holes with $\mathrm{He}=50$ in Figs.~\ref{fig:AWH}(a) and \ref{fig:AWH}(c).}
\label{fig:horizon_simAWH}
\end{figure}

Fig.~\ref{fig:AWH} shows the acoustic pressure distribution in space-time diagrams alongside the instantaneous wave profiles for the three acoustic white holes corresponding to the cases discussed in the previous section. The acoustic white hole can be seen as the inverse problem of the acoustic black hole \citep{Mayoral_et_al_2011}, where the flow field, still obeying Eq.~\eqref{eq:u0_spherical}, decelerates so that the ingoing characteristics cannot enter the white hole region from outside the sonic horizon. The physical domain $\Omega\left(t\right)$ is initially fully located inside the white hole, and, subsequently, the wave-emitting boundary expands and passes the sonic horizon, where again the positive peak of the eighth emitted period coincides with \revtwo{the sonic horizon at} $r_{\mathrm{h}}$. As a consequence, neighboring wave characteristics converge, which is accompanied by an increase of the acoustic pressure, as observed in Fig.~\ref{fig:horizon_simAWH}. 
While this increase of the acoustic pressure along the sonic horizon is monotone for the linear waves, the pressure of the nonlinear waves increases initially but decreases after a sufficiently long time period as the distortion of the nonlinear wave becomes eventually dominant. \revall{It is important to remember that this difference between the linear and nonlinear cases is due to the wave being distorted and the pressure amplitude recorded at the sonic horizon changing as a result, as clearly visible in the instantaneous wave profiles shown in Fig.~\ref{fig:AWH}.}
The $\mathrm{He}$-similarity of the wave characteristics and the pressure observed for the acoustic black hole in Sec.~\ref{sec:Acoustic black hole configuration} also holds for the acoustic white hole configuration, for both the linear and nonlinear waves.

\section{Leading-order model}\label{sec:Local analytical model}

The results presented in the previous sections suggest that the modulation of the acoustic pressure observed in the acoustic black and white hole configurations is caused by the divergence/convergence of the wave characteristics, which, in turn, is driven by the accelerating background flow. In the following, a reduced-order model is proposed to obtain further insight into the physical mechanisms underlying the observed modulation and similarity of acoustic waves in accelerating flows.

\begin{figure}
    {\includegraphics[width=0.95\reprintcolumnwidth]{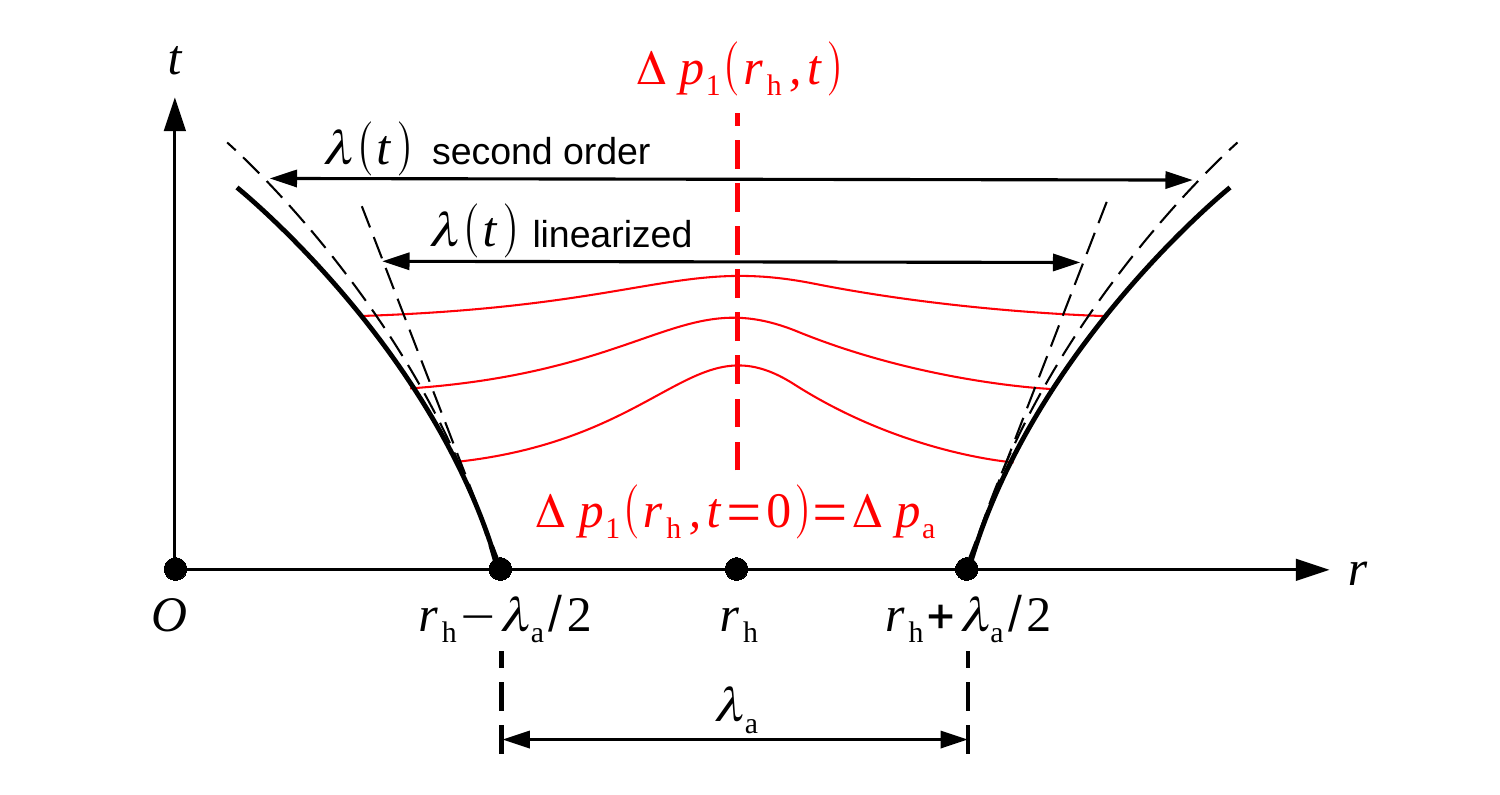}}
    \caption{Schematic of the modulation of the acoustic waves due to the divergence of the wave characteristics neighboring the sonic horizon.}
    \label{fig:romodel}
\end{figure}

In the considered acoustic black and white hole configurations, each acoustic wavelet of length $\lambda$ is endowed with a certain acoustic energy upon emission. By virtue of energy conservation, established in the present modeling framework by the transient Bernoulli equation, Eq.~\eqref{eq:Bernoulli}, it is assumed that the acoustic energy of a wavelet in a non-uniform background flow, given by \citep{Rienstra2004}
\begin{equation}
E = \int_{\lambda}\left(\dfrac{p_1^2}{2\rho_0c_0^2} + \dfrac{\rho_0 u_1^2}{2} + u_0\rho_1u_1\right) \, \mathrm{d}V,
\label{eq:soundEnergy}
\end{equation}
is conserved as $\lambda$ changes in the accelerating background flow, as illustrated in Fig.~\ref{fig:romodel}. Employing the linear acoustic relation $p_1=\rho_0c_0 u_1$, Eq.~\eqref{eq:soundEnergy} simplifies to
\begin{equation}
E = \dfrac{1}{\rho_0c_0^2}\int_{\lambda} \left(1 + \dfrac{u_0}{c_0}\right)p_1^2 \, \mathrm{d}V.
\label{eq:soundEnergy2}
\end{equation}
A linearized quasi one-dimensional approximation of the acoustic energy of each wavelet, Eq.~\eqref{eq:soundEnergy2}, is given by
\begin{equation}
E \approx \dfrac{\Delta A}{\rho_0c_0^2}\left(1 + \dfrac{\Delta u_0}{c_0}\right)\int_{\lambda}p_1^2 \, \mathrm{d}r \overset{|\Delta u_0| \ll c_0}{\approx}  \dfrac{\Delta A}{\rho_0c_0^2}\int_{\lambda}p_1^2 \, \mathrm{d}r, 
\label{eq:soundEnergy2_q1d}
\end{equation}
where $\Delta A$ denotes the change of cross-sectional area over the wavelet, approximated by linearization around the sonic horizon as $\Delta A = (\partial A/\partial r)_{r_{\mathrm{h}}}\lambda$. Likewise, $\Delta u_0$ is the change of \revtwo{the background flow velocity} $u_0$ over the wavelet, a contribution that can be neglected since $|\Delta u_0/c_0|\ll 1$.

Assuming a locally sinusoidal wavelet with instantaneous acoustic pressure amplitude $\Delta p_1$ and wavenumber $k$, the integral in Eq.~\eqref{eq:soundEnergy2_q1d} follows as
\begin{equation}
\int_{\lambda}p_1^2 \mathrm{d}r = \int_0^{\lambda}  \Delta {p}_1^2 \sin^2\left(kr\right) \mathrm{d}r = \dfrac{\lambda \Delta p_1^2}{2}.
\label{eq:int_p1_square}
\end{equation}
With Eq.~\eqref{eq:int_p1_square} and the relation $\Delta A \propto \lambda$, Eq.~\eqref{eq:soundEnergy2_q1d} becomes $E \propto \lambda^2 \Delta {p}_1^2$, where the initial condition $E= \mathcal{C} \lambda_{\mathrm{a}}^2\Delta p_{\mathrm{a}}^2$ with constant $\mathcal{C}$ yields \revtwo{the acoustic pressure amplitude defined as}
\begin{equation}
\Delta {p}_1(t) = \Delta p_{\mathrm{a}}\dfrac{\lambda_{\mathrm{a}}}{\lambda(t)} ,
\label{eq:soundEnergy3}
\end{equation}
\revtwo{where $\Delta p_\mathrm{a}$ and $\lambda_\mathrm{a}$ denote the excitation pressure amplitude and excitation wavelength of the acoustic waves, respectively.}
The time-dependent wavelength $\lambda(t)$ at the sonic horizon is obtained by linearizing the background flow, 
\begin{equation}
\lambda (t) = \lambda_{\mathrm{a}}  \pm \left.\dfrac{\partial u_0}{\partial r}\right|_{r_{\mathrm{h}}}\lambda_{\mathrm{a}}t 
=  \lambda_\mathrm{a} + \frac{a_\mathrm{h}}{c_0} \, \lambda_\mathrm{a} t.
\label{eq:linLambda}
\end{equation}
Note that \revtwo{the acceleration, $a_\mathrm{h}$, of the background flow at the sonic horizon} is by definition positive in the direction of the background flow (see Sec.~\ref{sec:Setup}). 
Introducing the modified Helmholtz number \revtwo{$\mathrm{He}$,} Eq.~\eqref{eq:HelmholtzABH}, and using the dispersion relation $c_0=\lambda_{\mathrm{a}}f_{\mathrm{a}}$, \revtwo{where $f_\mathrm{a}$ is the excitation frequency of the acoustic waves,} the time-dependent wavelength at the sonic horizon is
\begin{equation}
\lambda (t) = \lambda_\mathrm{a} \left(1 \pm \frac{f_\mathrm{a} t}{\mathrm{He}} \right) = \lambda_\mathrm{a} \left(1 + \frac{a_\mathrm{h} t}{c_0} \right). \label{eq:linLambda2}
\end{equation}
Substituting Eq.~\eqref{eq:linLambda2} into Eq.~\eqref{eq:soundEnergy3} gives the relation
\begin{equation}
\Delta {p}_1(t) = \Delta p_{\mathrm{a}} \, \dfrac{1}{1 \pm \dfrac{f_{\mathrm{a}}t}{\mathrm{He}}} = \Delta p_{\mathrm{a}} \, \dfrac{1}{1 + \dfrac{a_\mathrm{h} t}{c_0}}.
\label{eq:phat}
\end{equation}

\begin{figure}
    {\includegraphics[width=\reprintcolumnwidth]{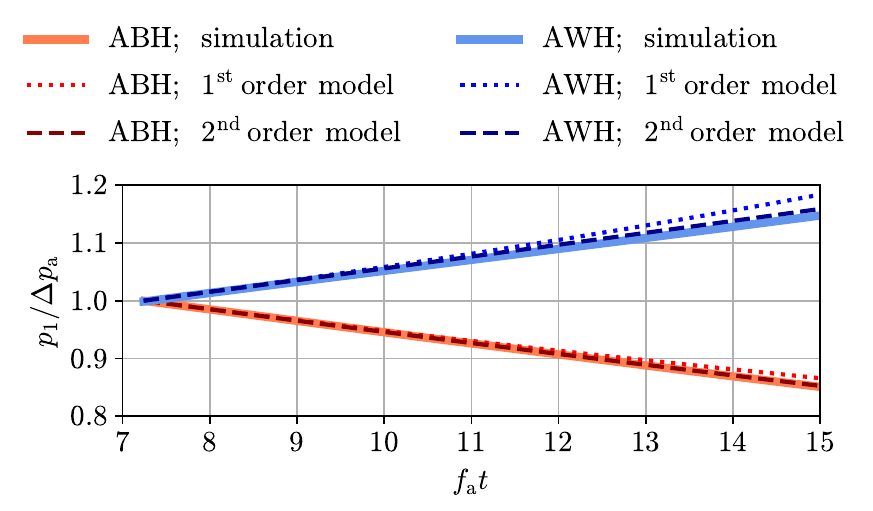}}
    \caption{Evolution of the acoustic pressure $p_1$ at the sonic horizon $r_{\mathrm{h}}$ for linear waves in acoustic black hole (ABH) and  white hole (AWH) configurations with $\mathrm{He}=50$. The first-order solution is given by Eq.~\eqref{eq:phat}, and the second-order solution by Eq.~\eqref{eq:phatO2}.}
    \label{fig:horizon_ABH_AWH}
\end{figure}

As observed in Fig.~\ref{fig:horizon_ABH_AWH}, this leading-order model, with the pressure amplitude $\Delta {p}_1(t)$ given by Eq.~\eqref{eq:phat}, describes the amplitude modulation in both the acoustic black and white hole configurations with very good accuracy for sufficiently short times $t$. The comparison between the numerical and the analytical result is facilitated by calibrating the initial wave-emitting boundary position $R(t=0)$ in such a way that, for the linear wave, the acoustic pressure $p_1$ at \revtwo{the sonic horizon} $r_{\mathrm{h}}$ coincides with the wave amplitude $\Delta p_1$ throughout the entire simulation.
The progressive departure from the numerical result is expected, since the leading-order model fails to capture the curvature of the wave characteristics. Expanding this model to higher order using Taylor series expansions, see Appx.~\ref{sec:Extrapolation to second order}, provides an even closer agreement with the numerical results, as also seen in Fig.~\ref{fig:horizon_ABH_AWH}, supporting the employed assumptions.

Eqs.~\eqref{eq:linLambda2} and \eqref{eq:phat} demonstrate that the modified Helmholtz number $\mathrm{He}$ and the dimensionless time $f_\mathrm{a}t$ jointly provide a suitable parameterization of acoustic waves in accelerating flows. For the acoustic energy of a wavelet to be conserved, its amplitude reduces if the wave characteristics are diverging as a result of an accelerating background flow, and its amplitude increases if the wave characteristics are converging \revall{as a result of a decelerating background flow}. 

The observed modulations of the wavelength and pressure amplitude are distinct from the conventional Doppler effect and the convective amplification associated with a moving emitter. While both Eqs.~\eqref{eq:linLambda2} and \eqref{eq:phat} depend on what can be interpreted as a Doppler factor $(1 + M^\star)^{-1}$, where $M^\star =  a_\mathrm{h} t/c_0$ is a characteristic Mach number, this Doppler factor depends on the observation time $t$. By contrast, the change in wavelength (or frequency) of the conventional Doppler effect depends on the Mach number of the relative velocity between emitter and observer, whereas the convective amplification of the pressure amplitude depends on the Mach number associated with the velocity of the emitter relative to the surrounding fluid \citep{Crighton1992}. The difference is that, here, the relative velocity between the emitter and the background fluid vanishes. Thus, the modulation of the wavelength and, consequently, the pressure amplitude are driven exclusively by the acceleration of the background flow, explaining why the Doppler factor $(1 + M^\star)^{-1}$ in Eq.~\eqref{eq:phat} depends on the observation time.

It is further noted that, to leading order, the modulation of the acoustic pressure amplitude $\Delta {p}_1(t)$, Eq.~\eqref{eq:phat}, is independent of the initial wavelength $\lambda_\mathrm{a}$ or frequency $f_\mathrm{a}$, and only depends on the observation time $t$, the acceleration $a_\mathrm{h}$, and the speed of sound $c_0$ of the background fluid. Hence, applying the notion of a virtual acoustic black/white hole introduced in Sec.~\ref{sec:Preliminary considerations}, this leading-order model may be used to predict or analyze acceleration-driven amplitude modulations, by numerically integrating the evolution of the wave characteristics, Eq.~\eqref{eq:characteristic}, and the associated amplitude modulation, Eq.~\eqref{eq:phat}.

\section{Conclusions}\label{sec:Conclusion}

Isolating and distinguishing the physical mechanisms modulating acoustic waves in moving and accelerating flows are important for a more detailed understanding and precise interpretation of the information encoded in acoustic signals. In this article, we have proposed a model system based on acoustic black and white hole analogues that facilitate the investigation of the modulation of acoustic waves in accelerating flows. To this end, we have presented a numerical framework based on a convective Kuznetsov equation that accounts for the background flow and a moving wave-emitting boundary, which remains well posed for transonic flows and is solved numerically using a finite-difference method.

The results presented in this paper identify a modulation of the wave amplitude caused by the  divergence or convergence of the wave characteristics, i.e.,~the motion of the wave characteristics relative to each other in the space-time plane, which corresponds to a expansion or compression of the acoustic waves. At the sonic horizon of the considered acoustic black/white hole configurations, the divergence/convergence of the wave characteristics is driven exclusively by the acceleration/deceleration of the background flow. Both the change in wavelength as well as the ensuing modulation of the acoustic pressure amplitude are self similar with respect to a modified Helmholtz number. We have exploited this similarity to derive a leading-order model based on first principles that provides further insight into the physical mechanisms underlying the observed wave modulation in accelerating flows and that can approximate the amplitude modulation over short time intervals accurately.

Considering a Galilean transformation in space-time, the employed acoustic black and white hole analogues can be applied to acoustic waves in any accelerating flow, assuming the characteristic defining the change of reference frame has a locally vanishing curvature. Thus, the proposed model system and the accompanying numerical framework may serve as a test bed for the study of the complex behavior of acoustic waves in accelerating flows. Moreover, the presented leading-order model can be used to predict the amplitude modulation of acoustic waves in accelerating flows, as well as to quantify the influence of the flow acceleration on acoustic waves in existing numerical simulation results and experimental measurements.


\section*{Acknowledgements}
We thank Mohammad Rashik Niaz and Rishav Saha for their help with some implementation details and with testing of the employed software tool. This research was funded by the Deutsche Forschungsgemeinschaft (DFG, German Research Foundation), grant number 441063377.

\section*{Declaration of competing interest}
The authors declare that they have no known competing financial interests or personal relationships that could have appeared to influence the work reported in this paper.

\appendix

\section{Higher-order model}\label{sec:Extrapolation to second order}

The leading-order model (Sec.~\ref{sec:Local analytical model}) is extended to higher order using Taylor series expansions. For brevity, the prime denotes the spatial derivative $\partial/\partial r$. First, the expansion of $A\left(r\right)$ around $r_{\mathrm{h}}$ with respect to $\lambda$ yields $\Delta A = A'\lambda + A'''\lambda^3/24$. Eq.~\eqref{eq:soundEnergy2_q1d} then yields
\begin{equation}
\Delta p_1(t) = \Delta p_{\mathrm{a}} \ \sqrt{\dfrac{A'\lambda_{\mathrm{a}}^2 + A'''\lambda_{\mathrm{a}}^4/24}{A'\lambda(t)^2 + A'''\lambda(t)^4/24}},
\label{eq:phatO2}
\end{equation}
where, analogous to the leading-order model, $\Delta p_{\mathrm{a}}$ and $\lambda_{\mathrm{a}}$ are introduced by requiring the initial condition to satisfy $E=\mathrm{const}$. Next, the linear relation for $\lambda(t)$ given by Eq.~\eqref{eq:linLambda} is expanded in both space and time, where $v_{\mathrm{div}}(t) = \left(\partial u_0/\partial r\right)_{r_{\mathrm{h}}}\lambda(t)$ can be interpreted as the speed at which the length of the considered wavelet is changing, and which varies as the wave characteristic evolves in the spatially non-uniform flow field. Expanding the expression $\lambda\left(t\right) = \lambda_{\mathrm{a}} \pm v_{\mathrm{div}}(t) \ t$ to second order gives
\begin{equation}
\lambda \left(t\right) = \lambda_{\mathrm{a}}  \pm v_{\mathrm{div}}(t) \ t \pm \dfrac{1}{2}\left.\dfrac{\mathrm{D}v_{\mathrm{div}}(t)}{\mathrm{D}t}\right|_{r_{\mathrm{h}}}t^2,
\label{eq:lambdaO2}
\end{equation}
where the temporal change of $v_{\mathrm{div}}$ along the characteristics depends on the material derivative. Finally, $v_{\mathrm{div}}$ is expanded in space, analogously to the expansion of $\Delta A$, yielding $v_{\mathrm{div}}(t) = u_0'\lambda(t) + u_0'''\lambda(t)^3/24$. Evaluating this expression at the initial time, where $\lambda = \lambda_{\mathrm{a}}$, inserting into Eq.~\eqref{eq:lambdaO2} and taking into account that $\mathrm{D}/\mathrm{D}t = u_0\partial/\partial r$ for the stationary flow field in the considered acoustic black and white hole configurations, gives
\begin{equation}
\lambda \left(t\right) = \lambda_{\mathrm{a}}  
\pm 
\left[u_0'\lambda_{\mathrm{a}} + \dfrac{u_0'''}{24}\lambda_{\mathrm{a}}^3\right]_{r_{\mathrm{h}}} t 
\pm 
\dfrac{u_0}{2}\left[u_0''\lambda_{\mathrm{a}} + \dfrac{u_0''''}{24}\lambda_{\mathrm{a}}^3\right]_{r_{\mathrm{h}}} t^2.
\label{eq:lambdaO2_2}
\end{equation}
Eq.~\eqref{eq:phatO2}, in conjunction with Eq.~\eqref{eq:lambdaO2_2}, represents a second-order approximation of the modulation of an acoustic wave in an accelerating background flow. For the spherical flow considered here, $A'=2/r$, $A''=-2/r^2$, and $A'''=4/r^3$. The spatial derivatives of $u_0$, evaluated at $r_{\mathrm{h}}$, follow as $u_0'=\pm 2c_0/r_{\mathrm{h}}$, $u_0''=\mp 6c_0/r_{\mathrm{h}}^2$, $u_0'''=\pm 24c_0/r_{\mathrm{h}}^3$, and $u_0''''=\mp 120c_0/r_{\mathrm{h}}^4$ for the black and white hole configurations, respectively.

\newcommand{\noop}[1]{}

\end{document}


\pagestyle{fancy}
\noindent 
\begin{center}
\textbf{\Large Amplitude modulation of acoustic waves in accelerating flows\\[-0.2em] quantified using acoustic black and white hole analogues\\[0.5em] \underline{\large -- Supplementary material --}}\\[0.7em]
\noindent \textbf{S\"oren Schenke$^1$, Fabian Sewerin$^1$, Berend van Wachem$^1$, Fabian Denner$^{2,\dagger}$}\\[0.1em]
\noindent\textit{\small $^1$Chair of Mechanical Process Engineering, Otto-von-Guericke-Universit\"{a}t Magdeburg, Universit\"atsplatz 2, 39106 Magdeburg, Germany\\
$^2$Department of Mechanical Engineering, Polytechnique Montr\'eal, Montr\'eal, H3T 1J4, QC, Canada\\
$^\dagger$fabian.denner@polymtl.ca}\\ {\small \today}
\end{center}
\thispagestyle{empty}

Expanding on the manuscript, this supplementary material presents a detailed derivation and discussion of the proposed mathematical model and its numerical discretization. In the interest of completeness, we first revisit the modeling assumptions and conventions underpinning the proposed mathematical model in Section \ref{sec:assumptions}, as also stated in Section III.A of the manuscript. Based on these assumptions, Section \ref{sec:waveequation} presents a detailed step-by-step derivation of the lossless convective Kuznetsov equation proposed in Section III.B of the manuscript. The coordinate transform of the lossless convective Kuznetsov equation from the moving physical domain to the considered fixed computational domain, as introduced in Section III.C of the manuscript, is described in Section \ref{sec:coordinatetransformation}. Section \ref{sec:fd} is concerned with the finite-difference discretization of the transformed lossless convective Kuznetsov equation in the fixed computational domain, and lastly, Section \ref{sec:Transformation between the acoustic pressure and the perturbation potential} presents the transformation between the acoustic perturbation potential and the acoustic pressure field.
The presented modeling framework is implemented in the open-source software program {\tt Wave-DNA}, available at \url{https://doi.org/10.5281/zenodo.8084168}.

\section{Modeling assumptions and conventions}\label{sec:assumptions}

The governing equations are formulated for an inviscid quasi one-dimensional and, hence, irrotational flow with spatial coordinate $r$. The flow is assumed to pass a spatially variable cross-sectional area $A\left(r\right)$, with the Laplace operator given by
\begin{equation}
\nabla_{\mathrm{A}}^2 =
\dfrac{\partial^2}{\partial r^2} + \dfrac{1}{A}\dfrac{\partial A}{\partial
r}\dfrac{\partial}{\partial r}.
\label{eq:LaplacianDef}
\end{equation}
The continuity equation for this flow reads as 
\begin{equation}
\dfrac{\partial \rho}{\partial t} + \dfrac{1}{A}\dfrac{\partial}{\partial r}\left(A\rho u\right) = 0,
\label{eq:Conti}
\end{equation}
where $\rho$ and $u$ are the density and velocity of the fluid, and where $u$ follows from the velocity potential $\phi$ as $u = \partial \phi/\partial r$. The velocity potential $\phi$, the velocity $u$, the pressure $p$, and the density $\rho$ are formally decomposed into background contributions and acoustic perturbations by writing them in terms of the first-order series $\phi = \phi_0 + \epsilon\phi_1$, $u = u_0 + \epsilon u_1$, $p = p_0 + \epsilon p_1$, and $\rho = \rho_0 + \epsilon\rho_1$, where the subscript $0$ indicates the background state, which is assumed to be known, and the subscript $1$ indicates the unknown acoustic perturbation \citep{Visser_1998}. The dimensionless quantity $\epsilon$ is constant and equal to unity, and it is formally introduced to indicate the order of nonlinearity of the acoustic terms in the course of the following derivations. For brevity, $\epsilon$ is not written out in the further course of the derivation. It is noted, however, that terms involving the multiplication of any two acoustic terms with subscript $1$ formally represent nonlinear acoustic terms of $\mathcal{O}\left(\epsilon^2\right)$, which are consistently retained together with the first-order terms, whereas terms of $\mathcal{O}\left(\epsilon^{n\ge 3}\right)$ are neglected.  

{It is assumed that the fluid is barotropic, with $\mathrm{d} p/\mathrm{d} \rho = c_0^2 \approx p_1/\rho_1$, and that all compressibility effects are captured by the acoustic perturbations. Consequently, the background flow is assumed to be incompressible, with a constant density $\rho_0$ and speed of sound $c_0$. The background flow field $u_0\left(r,t\right)$, however, may be both spatially non-uniform and time-dependent. The material derivative operator is defined by the background flow field,}
\begin{equation}
\dfrac{\mathrm{D}}{\mathrm{D}t} = \dfrac{\partial}{\partial t} + u_0 \dfrac{\partial}{\partial r},
\label{eq:MaterialDerivative}
\end{equation}
and the Lagrangian density is given by \citep{Cervenka_and_Bednarik_2022}
\begin{equation}
\mathcal{L} = \dfrac{1}{2}\rho_0u_1^2 - \dfrac{1}{2}\dfrac{p_1^2}{\rho_0 c_0^2}
\label{eq:LagrangianDensity}.
\end{equation}

As discussed by \citet{Cervenka_and_Bednarik_2022}, the unknown nonlinear relation between $p_1$ and $u_1$ requires to formulate the second-order acoustic wave equation, i.e.,~the Kutznetsov equation, in terms of the acoustic potential $\phi_1$, where the relation $u_1 = \partial\phi_1/\partial r$ can be employed. The acoustic pressure $p_1$ is then obtained by consulting the conservation of energy or momentum. 

\section{Derivation of the lossless convective Kuznetsov equation}\label{sec:waveequation}

Following the suggestion of \citet{Visser_1998}, the potential form of the transient Bernoulli equation \citep{Prosperetti_1984},
\begin{equation}
\dfrac{\partial \phi}{\partial t} + \dfrac{1}{2}\left(\dfrac{\partial \phi}{\partial r}\right)^2 + h\left(p\right) = 0,
\label{eq:Bernoulli}
\end{equation} 
where $h\left(p\right) = \int_{p_0}^p \rho^{-1} \text{d}p$ is the enthalpy difference between $p$ and $p_0$, is used as the starting point for the derivation of the lossless convective Kuznetsov equation. Substituting the perturbations introduced in Section \ref{sec:assumptions} into Eqs.~\eqref{eq:Conti} and \eqref{eq:Bernoulli}, and isolating the acoustic terms of order $\mathcal{O}(\epsilon^1)$ and $\mathcal{O}(\epsilon^2)$, the conservation equations for the acoustic perturbations follow as
\begin{align}
\dfrac{\partial \rho_1}{\partial t} + \dfrac{1}{A}\dfrac{\partial}{\partial r}\left[A\left(\rho_0 u_1 + \rho_1 u_0 + \rho_1 u_1\right)\right] &= 0,
\label{eq:Conti1} \\
\dfrac{\mathrm{D} \phi_1}{\mathrm{D} t} + \dfrac{1}{2}\left(\dfrac{\partial \phi_1}{\partial r}\right)^2 +
h_1 &= 0,
\label{eq:Bernoulli_interim}
\end{align}
where $\partial \phi_0/\partial r = u_0$ per definition. The corresponding set of equations for the background flow fields is given by
\begin{align}
\dfrac{\partial u_0}{\partial r} + \dfrac{u_0}{A}\dfrac{\partial A}{\partial r} &= 0,
\label{eq:Conti_background} \\
\dfrac{\partial \phi_0}{\partial t} + \dfrac{1}{2}u_0^2 +
h_0 &= 0.
\label{eq:Bernoulli_background}
\end{align}
While the linearized acoustic perturbation enthalpy $h_1=p_1/\rho_0$ is used in the first-order theory of \citet{Unruh_1981} and \citet{Visser_1998}, this term must be expanded to second order in the context of the nonlinear theory in the present work. With the quadratic expansion \citep{Prosperetti_1984} 
\begin{equation}
h_1 = \dfrac{p_1}{\rho_0} - \dfrac{1}{2c_0^2}\left(\dfrac{p_1}{\rho_0}\right)^2,
\label{eq:hexpansion}    
\end{equation}
and substituting the first-order relation $p_1=-\rho_0\mathrm{D}\phi_1/\mathrm{D}t$ for the acoustic pressure $p_1$ in the quadratic term of Eq.~\eqref{eq:hexpansion},
Eq.~\eqref{eq:Bernoulli_interim} can be rewritten as
\begin{equation}
\dfrac{\mathrm{D}\phi_1}{\mathrm{D}t} +
\dfrac{\mathcal{L}}{\rho_0} +
\dfrac{p_1}{\rho_0} = 0,
\label{eq:Bernoulli1}
\end{equation}
with
\begin{equation}
\dfrac{\mathcal{L}}{\rho_0} = \dfrac{1}{2}\left(\dfrac{\partial\phi_1}{\partial r}\right)^2 - \dfrac{1}{2c_0^2}\left(\dfrac{\mathrm{D}\phi_1}{\mathrm{D} t}\right)^2.
\label{eq:LagrangianDensity2}
\end{equation}
Eq.~\eqref{eq:Bernoulli1} describes the mutual interaction between the acoustic pressure and the acoustic (particle) velocity, required to convert between the acoustic velocity potential $\phi_1$ and the acoustic pressure $p_1$. With the barotropic assumption, $\rho_1 = p_1/c_0^2$, and using the relation $\partial u_1/\partial r = \partial^2\phi_1/\partial r^2$, Eq.~\eqref{eq:Conti1} is rewritten as
\begin{align}
\dfrac{\partial \rho_1}{\partial t} + \dfrac{1}{A}\dfrac{\partial A}{\partial r}\left(\rho_0\dfrac{\partial \phi_1}{\partial r} + \dfrac{p_1}{c_0^2} \dfrac{\partial \phi_1}{\partial r}\right) + \rho_0\dfrac{\partial^2\phi_1}{\partial r^2} &+ \rho_1 \dfrac{\partial u_1}{\partial r}
 \nonumber \\
+ \dfrac{1}{c_0^2}\left(u_0\dfrac{\partial p_1}{\partial r} + \dfrac{\partial \phi_1}{\partial r}\dfrac{\partial p_1}{\partial r}\right) &= - \dfrac{p_1}{c_0^2} \underbrace{\left(\dfrac{u_0}{A}\dfrac{\partial A}{\partial r}  + \dfrac{\partial u_0}{\partial r}\right)}_{=0~\textnormal{(continuity equation)}},
\label{eq:Conti2}
\end{align}
Taking the continuity equation for the background flow field, Eq.~\eqref{eq:Conti_background}, into account, the right-hand side of Eq. \eqref{eq:Conti2} vanishes and the equation further simplifies to
\begin{equation}
\dfrac{\partial \rho_1}{\partial t} + \dfrac{1}{A}\dfrac{\partial A}{\partial r}\left(\rho_0\dfrac{\partial \phi_1}{\partial r} + \dfrac{p_1}{c_0^2} \dfrac{\partial \phi_1}{\partial r}\right) + \rho_0\dfrac{\partial^2\phi_1}{\partial r^2} + \rho_1 \dfrac{\partial u_1}{\partial r}
=
- \dfrac{1}{c_0^2}\left(u_0\dfrac{\partial p_1}{\partial r} + \dfrac{\partial \phi_1}{\partial r}\dfrac{\partial p_1}{\partial r}\right),
\label{eq:Conti3}
\end{equation}
Following the standard procedure of deriving second-order acoustic equations, the two factors $\rho_1$ and $\partial u_1/\partial r$ in the last term on the left-hand side of Eq. \eqref{eq:Conti3} are replaced by the first-order expressions of Eqs. \eqref{eq:Bernoulli1} and Eq. \eqref{eq:Conti3}, respectively, where the second order is preserved due to the multiplication of two first-order terms. With the barotropic assumption, the corresponding first-order expressions are
\begin{align}
& \rho_1 = - \dfrac{\rho_0}{c_0^2}\dfrac{\mathrm{D}\phi_1}{\mathrm{Dt}},
\label{eq:firstOrderBer} \\
& \dfrac{\partial u_1}{\partial r} = -\dfrac{1}{\rho_0}\dfrac{\partial \rho_1}{\partial t} - \dfrac{1}{A}\dfrac{\partial A}{\partial r}\dfrac{\partial \phi_1}{\partial r} - \dfrac{u_0}{\rho_0 c_0^2}\dfrac{\partial p_1}{\partial r},
\label{eq:firstOrderConti}
\end{align}
where Eq. \eqref{eq:firstOrderConti} is obtained by substituting the Laplacian of Eq. \eqref{eq:Conti3} by the expression $\partial^2\phi_1/\partial r ^2=\partial u_1/\partial r$ and subsequently solving for $\partial u_1/\partial r$, keeping the first-order terms only. Substituting Eqs. \eqref{eq:firstOrderBer} and \eqref{eq:firstOrderConti} into the last term on the left-hand side of Eq. \eqref{eq:Conti3} gives
\begin{equation}
\left(1 + \dfrac{1}{c_0^2}\dfrac{\mathrm{D}\phi_1}{\mathrm{D}t}\right)\dfrac{\partial \rho_1}{\partial t} + \rho_0\nabla_{\mathrm{A}}^2\phi_1
+ \dfrac{1}{c_0^2A}\dfrac{\partial A}{\partial r} \underbrace{\dfrac{\partial \phi_1}{\partial r}\left(p_1 + \rho_0\dfrac{\mathrm{D}\phi_1}{\mathrm{D}t}\right)}_{\approx0~\textnormal{(up to second order)}}
=
- \dfrac{1}{c_0^2}\left(u_0 + \dfrac{\partial \phi_1}{\partial r} + \dfrac{u_0}{c_0^2}\dfrac{\mathrm{D}\phi_1}{\mathrm{D}t}\right)\dfrac{\partial p_1}{\partial r},
\label{eq:Conti4}
\end{equation}
where $\nabla_{\mathrm{A}}$ is the Laplace operator as introduced in Eq. \eqref{eq:LaplacianDef}. It follows from Eq. \eqref{eq:Bernoulli1} that $p_1 + \rho_0\mathrm{D}\phi_1/\mathrm{D}t = -\mathcal{L}$ is a second-order term, which becomes a third-order term upon multiplication with $\partial\phi_1/\partial r$ and which can, therefore, be neglected. Eq. \eqref{eq:Conti4} then becomes
\begin{equation}
\left(1 + \dfrac{1}{c_0^2}\dfrac{\mathrm{D}\phi_1}{\mathrm{D}t}\right)\dfrac{\partial \rho_1}{\partial t} + \rho_0\nabla_{\mathrm{A}}^2\phi_1
=
- \dfrac{1}{c_0^2}\left(u_0 + \dfrac{\partial \phi_1}{\partial r} + \dfrac{u_0}{c_0^2}\dfrac{\mathrm{D}\phi_1}{\mathrm{D}t}\right)\dfrac{\partial p_1}{\partial r}.
\label{eq:Conti5}
\end{equation}
which, after rearranging, can be written as
\begin{equation}
\left(1+ \dfrac{1}{c_0^2}\dfrac{\mathrm{D}\phi_1}{\mathrm{D}t}\right)\left(\dfrac{\partial \rho_1}{\partial t} + \dfrac{u_0}{c_0^2}\dfrac{\partial p_1}{\partial r}\right) + \rho_0 \nabla_{\mathrm{A}}^2\phi_1 + \dfrac{1}{c_0^2}\dfrac{\partial \phi_1}{\partial r}\dfrac{\partial p_1}{\partial r} = 0.
\label{eq:Conti5b}
\end{equation}
With the relation $\rho_1=p_1/c_0^2$ and the definition of the material derivative operator as given by Eq. \eqref{eq:MaterialDerivative}, and by multiplying with $c_0^2$, Eq.~\eqref{eq:Conti5b} can be rewritten in the more compact form
\begin{equation}
\left(1+ \dfrac{1}{c_0^2}\dfrac{\mathrm{D}\phi_1}{\mathrm{D}t}\right)\dfrac{\mathrm{D}p_1}{\mathrm{D}t} + \rho_0 c_0^2\nabla_{\mathrm{A}}^2\phi_1 + \dfrac{1}{c_0^2}\dfrac{\partial \phi_1}{\partial r}\dfrac{\partial p_1}{\partial r} = 0.
\label{eq:Conti6}
\end{equation}
The acoustic pressure state is expanded to second order such that \citep{Shevchenko_and_Kaltenbacher_2015}
\begin{equation}
\dfrac{p_1}{c_0^2} = \rho_1 - \dfrac{B}{2A}\dfrac{\rho_1^2}{\rho_0},
\label{eq:pexpansion}
\end{equation}
where $B/(2A)$ is the nonlinearity parameter, specifying a medium-dependent weight of the nonlinear acoustic contribution relative to the linear acoustic contribution. Note that this step is similar to the quadratic expansion of the linearized perturbation enthalpy $h_1$ in Eq.~\eqref{eq:hexpansion}. With the nonlinearity coefficient $\beta = 1 + B/(2A)$ and substituting $\rho_1 = p_1/c_0^2$ by Eq.~\eqref{eq:Bernoulli1}, we obtain 
\begin{equation}
\dfrac{p_1}{\rho_0} = c_0^2\dfrac{\rho_1}{\rho_0} = - \dfrac{\mathrm{D}\phi_1}{\mathrm{D}t} - \dfrac{\mathcal{L}}{\rho_0} - \dfrac{\beta - 1}{c_0^2}\left(\dfrac{\mathrm{D}\phi_1}{\mathrm{D}t}\right)^2.
\label{eq:rhoExpansion}
\end{equation}
From on Eq. \eqref{eq:rhoExpansion}, the derivatives $\mathrm{D} p_1/\mathrm{D} t$ and $\partial p_1/\partial r$ of the expanded pressure state follow as
\begin{align}
& \dfrac{\mathrm{D}p_1}{\mathrm{D}t} = -\rho_0\dfrac{\mathrm{D}^2\phi_1}{\mathrm{D}t^2} - \dfrac{\mathrm{D}\mathcal{L}}{\mathrm{D}t} - 2\rho_0\dfrac{\beta - 1}{c_0^2}\dfrac{\mathrm{D}\phi_1}{\mathrm{D}t}\dfrac{\mathrm{D}^2\phi_1}{\mathrm{D}t^2},
\label{eq:expansionDDt} \\
& \dfrac{\partial p_1}{\partial r} = -\rho_0\dfrac{\partial}{\partial r}\left(\dfrac{\mathrm{D}\phi_1}{\mathrm{D}t}\right) - \dfrac{\partial \mathcal{L}}{\partial r} - 2\rho_0\dfrac{\beta - 1}{c_0^2}\dfrac{\mathrm{D}\phi_1}{\mathrm{D}t}\dfrac{\partial}{\partial r}\left(\dfrac{\mathrm{D}\phi_1}{\mathrm{D}t}\right).
\label{eq:expansionddr}
\end{align}
Substituting Eqs. \eqref{eq:expansionDDt} and \eqref{eq:expansionDDt} into Eq. \eqref{eq:Conti6}, keeping only terms up to second order, $\mathcal{O}\left(\epsilon^2\right)$, and dividing by $\rho_0$ yields
\begin{equation}
\dfrac{\mathrm{D}^2\phi_1}{\mathrm{D}t^2} + 2\dfrac{\beta - 1}{c_0^2}\dfrac{\mathrm{D}\phi_1}{\mathrm{D}t}\dfrac{\mathrm{D}^2\phi_1}{\mathrm{D}t^2}
+ \dfrac{1}{\rho_0}\dfrac{\mathrm{D}\mathcal{L}}{\mathrm{D}t}
+\dfrac{1}{c_0^2}\dfrac{\mathrm{D}\phi_1}{\mathrm{D}t}\dfrac{\mathrm{D}^2\phi_1}{\mathrm{D}t^2}
- c_0^2\nabla_{\mathrm{A}}^2\phi_1
+ \dfrac{\partial \phi_1}{\partial r}\dfrac{\partial}{\partial r}\left(\dfrac{\mathrm{D}\phi_1}{\mathrm{D}t}\right) = 0.
\label{eq:Conti7}
\end{equation}
The terms in Eq. \eqref{eq:Conti7} can be summarized to yield the lossless convective Kuznetsov equation
\begin{equation}
\left(1 + \dfrac{2\beta - 1}{c_0^2}\dfrac{\mathrm{D}\phi_1}{\mathrm{D}t}\right)\dfrac{\mathrm{D}^2\phi_1}{\mathrm{D}t^2} 
- c_0^2\nabla_{\mathrm{A}}^2\phi_1
+ \dfrac{1}{\rho_0}\dfrac{\mathrm{D}\mathcal{L}}{\mathrm{D}t}
+ \dfrac{\partial \phi_1}{\partial r}\dfrac{\partial}{\partial r}\left(\dfrac{\mathrm{D}\phi_1}{\mathrm{D}t}\right) = 0,
\label{eq:convectiveKuznetsov}
\end{equation}
where the potential form of $\mathcal{L}$ is given by Eq.~\eqref{eq:LagrangianDensity2}.



In the limit of a vanishing background flow with $u_0\rightarrow 0$ and, consequently, $\mathrm{D}/\mathrm{D}t\rightarrow\partial/\partial t$, the lossless convective Kuznetsov equation as presented in Eq. \eqref{eq:convectiveKuznetsov} reduces to
\begin{equation}
\dfrac{\partial^2\phi_1}{\partial t^2} + \dfrac{2\beta - 1}{c_0^2}\dfrac{\partial\phi_1}{\partial t}\dfrac{\partial^2\phi_1}{\partial t^2} 
- c_0^2\nabla_{\mathrm{A}}^2\phi_1
+ \dfrac{1}{\rho_0}\dfrac{\partial \mathcal{L}}{\partial t}
+ \dfrac{\partial \phi_1}{\partial r}\dfrac{\partial^2\phi_1}{\partial r\partial t} = 0.
\label{eq:convectiveKuznetsov_interim}
\end{equation}
With
\begin{align}
&\dfrac{\partial}{\partial t}\left[\left(\dfrac{\partial\phi_1}{\partial t}\right)^2\right] = 2\dfrac{\partial \phi_1}{\partial t}\dfrac{\partial^2\phi_1}{\partial t^2},
\label{eq:squder_t}
\\
&\dfrac{\partial}{\partial t}\left[\left(\dfrac{\partial\phi_1}{\partial r}\right)^2\right] = 2\dfrac{\partial \phi_1}{\partial r}\dfrac{\partial^2\phi_1}{\partial r\partial t},
\label{eq:squder_r}
\end{align}
Eq.~\eqref{eq:convectiveKuznetsov_interim} can be recast into the form
\begin{equation}
c_0^2\nabla_{\mathrm{A}}^2\phi_1
-\dfrac{\partial^2\phi_1}{\partial t^2}
=
\dfrac{\partial}{\partial t}\left[\left(\dfrac{\partial\phi_1}{\partial r}\right)^2 + \dfrac{\beta-1}{c_0^2}\left(\dfrac{\partial\phi_1}{\partial t}\right)^2\right],
\label{eq:standardKuznetsov}
\end{equation}
which is the lossless form of the standard Kuznetsov equation \citep{Cervenka_and_Bednarik_2022}. 

In the linear limit, neglecting second-order nonlinear terms, the lossless convective Kuznetsov equation as presented in Eq. \eqref{eq:convectiveKuznetsov} reduces to the convective wave equation \citep{Kaltenbacher_and_Hueppe_2018}
\begin{equation}
\dfrac{\mathrm{D}^2\phi_1}{\mathrm{D}t^2} - c_0^2\nabla_{\mathrm{A}}^2\phi_1 = 0,
\label{eq:convectiveWaveEqn}
\end{equation}
which agrees with the wave equation presented by \citet{Unruh_1981} and \citet{Visser_1998} once the assumption of a constant background density $\rho_0$ and constant speed of sound $c_0$ is applied.

\section{Coordinate transformation}
\label{sec:coordinatetransformation}

The position of the wave-emitting domain boundary is assumed to be time dependent. Similar to the approaches by \citet{Christov_2017} and \citet{Gasperini_et_al_2021}, we invoke a time-dependent coordinate transformation \citep{Schenke_et_al_2022}
\begin{equation}
r: \: \left[\mathcal{X}_R,\mathcal{X}_{\infty}\right]
\times \left[0, \infty\right)\rightarrow \left[R\left(t\right),R_{\mathrm{stat}}\right], \: \left(\xi,t\right) \mapsto r\left(\xi,t\right)
\label{eq:coordTrans}
\end{equation}
between a moving physical domain $\Omega\left(t\right)$ with the time-dependent left boundary $R\left(t\right)$ and a fixed right boundary $R_{\mathrm{stat}}$, and a fixed computational domain $\Theta$ with spatial coordinate $\xi$ and fixed left and right boundaries $\mathcal{X}_R$ and $\mathcal{X}_{\infty}$, in which the transformed governing equation is solved by means of standard finite-difference techniques. Without loss of generality, the fixed domain boundaries of $\Theta$ are chosen to be $\mathcal{X}_R=0$ and $\mathcal{X}_{\infty}=1$. The coordinate transformation conveys a mapping between the two domains, such that $\phi_1\left(r\left(\xi,t\right),t\right)=\Phi_1\left(\xi,t\right)$. The $k$-th spatial and temporal derivatives of the acoustic potential $\phi_1$ can then be written as
\begin{align}
\dfrac{\partial^k \phi_1}{\partial r^k} &= \left(\dfrac{\partial \xi}{\partial r} \dfrac{\partial}{\partial \xi}\right)^k\Phi_1,
\label{eq:transDerivativeOp_space} \\
\dfrac{\partial^k \phi_1}{\partial t^k} &= \left(\dfrac{\partial \xi}{\partial t} \dfrac{\partial}{\partial \xi} + \dfrac{\partial}{\partial t}\right)^k\Phi_1.
\label{eq:transDerivativeOp_time}
\end{align}
The derivatives of $\xi$ depend on the employed coordinate transformation. We opt for the linear transformation
\begin{equation}
\xi\left(r,t\right) = \mathcal{X}_{\infty} + \left(r - R_{\mathrm{stat}}\right) \dfrac{\mathcal{X}_{\mathrm{\infty}} - \mathcal{X}_{R}}{R_{\mathrm{stat}}-R\left(t\right)}.
\label{eq:linTrans}
\end{equation}

To change the variable of Eq.~\eqref{eq:convectiveKuznetsov} from $\phi_1$ to $\Phi_1$, it is beneficial to first define the recurring auxiliary variables
\begin{align}
\mathcal{K} &= \dfrac{2\left(\beta - 1\right)}{c_0^2},
\label{eq:Kcoeff}
\\
\mathcal{U} &= 2u_0\dfrac{\partial^2 \phi_1}{\partial x\partial t} + \dfrac{\mathrm{D}u_0}{\mathrm{D}t}\dfrac{\partial\phi_1}{\partial r},
\label{eq:Uterm}
\end{align}
with which Eq. \eqref{eq:convectiveKuznetsov} is rewritten as
\begin{equation}
\left(1 + \mathcal{K} \, \dfrac{\mathrm{D}\phi_1}{\mathrm{D}t}\right)\dfrac{\mathrm{D}^2\phi_1}{\mathrm{D}t^2} 
- c_0^2\nabla_{\mathrm{A}}^2\phi_1
+ \dfrac{1}{\rho_0}\dfrac{\mathrm{D}\mathcal{L}}{\mathrm{D}t}
+ \dfrac{\partial \phi_1}{\partial r}\dfrac{\partial}{\partial r}\left(\dfrac{\mathrm{D}\phi_1}{\mathrm{D}t}\right) = 0.
\label{eq:convectiveKuznetsov2}
\end{equation}
Expanding the material derivatives of $\phi_1$ using Eq. \eqref{eq:MaterialDerivative} yields
\begin{align}
\dfrac{\mathrm{D}^2\phi_1}{\mathrm{D}t^2} &= \dfrac{\partial^2\phi_1}{\partial t^2} + 2u_0\dfrac{\partial^2\phi_1}{\partial r\partial t} + \dfrac{\mathrm{D}u_0}{\mathrm{D}t} \dfrac{\partial \phi_1}{\partial r} + u_0^2\dfrac{\partial^2\phi_1}{\partial r^2},
\label{eq:D2phi}
\\
\mathcal{K} \, \dfrac{\mathrm{D}\phi_1}{\mathrm{D}t}\dfrac{\mathrm{D}^2\phi_1}{\mathrm{D}t^2}
&=
\mathcal{K}\, u_0\dfrac{\partial \phi_1}{\partial r}\dfrac{\partial^2\phi_1}{\partial t^2} 
+ \mathcal{K}\left(2u_0\dfrac{\partial^2\phi_1}{\partial r\partial t} + \dfrac{\mathrm{D}u_0}{\mathrm{D}t} \dfrac{\partial \phi_1}{\partial r} + u_0^2\dfrac{\partial^2\phi_1}{\partial r^2} \right)\dfrac{\partial \phi_1}{\partial t} 
\nonumber
\\
& + \mathcal{K}\, \dfrac{\partial \phi_1}{\partial t}\dfrac{\partial^2 \phi_1}{\partial t^2}
+ \mathcal{K}\, u_0\left(2u_0\dfrac{\partial^2\phi_1}{\partial r\partial t} + \dfrac{\mathrm{D}u_0}{\mathrm{D}t} \dfrac{\partial \phi_1}{\partial r} + u_0^2\dfrac{\partial^2\phi_1}{\partial r^2} \right)\dfrac{\partial \phi_1}{\partial r}, \label{eq:KD1phiD2phi}
\\
\dfrac{\partial \phi_1}{\partial r}\dfrac{\partial}{\partial r}\left(\dfrac{\mathrm{D}\phi_1}{\mathrm{D}t}\right)
& = 
u_0\dfrac{\partial \phi_1}{\partial r}\dfrac{\partial^2\phi_1}{\partial r^2} + \left(\dfrac{\partial^2\phi_1}{\partial r\partial t} + \dfrac{\partial u_0}{\partial r} \dfrac{\partial \phi_1}{\partial r}\right)\dfrac{\partial \phi_1}{\partial r}.
\label{eq:gradphigradDphi}
\end{align}
The material derivative of the Lagrangian density $\mathcal{L}$ can be written as
\begin{equation}
\dfrac{1}{\rho_0}\dfrac{\mathrm{D}\mathcal{L}}{\mathrm{D}t}
= 
\left(\dfrac{\partial}{\partial t} + u_0\dfrac{\partial}{\partial r}\right)\left[\dfrac{1}{2}\left(\dfrac{\partial\phi_1}{\partial r}\right)^2 - \dfrac{1}{2c_0^2}\left(\dfrac{\mathrm{D}\phi_1}{\mathrm{D} t}\right)^2\right]
\label{eq:DLDt}
\end{equation}
Expanding the derivatives in Eq. \eqref{eq:DLDt} yields
\begin{align}
\dfrac{1}{\rho_0}\dfrac{\mathrm{D}\mathcal{L}}{\mathrm{D}t}
= &
\dfrac{\partial^2\phi_1}{\partial r\partial t}\dfrac{\partial \phi_1}{\partial r} + u_0\dfrac{\partial\phi_1}{\partial r}\dfrac{\partial^2\phi_1}{\partial r^2}
\nonumber
\\
- & \dfrac{1}{c_0^2}\dfrac{\partial\phi_1}{\partial t}\dfrac{\partial^2\phi_1}{\partial t^2}
- \dfrac{1}{c_0^2}\dfrac{\partial u_0}{\partial t}\dfrac{\partial\phi_1}{\partial r}\dfrac{\partial\phi_1}{\partial t}
- \dfrac{u_0}{c_0^2}\dfrac{\partial^2\phi_1}{\partial r\partial t}\dfrac{\partial\phi_1}{\partial t}
\nonumber
\\
- & \dfrac{u_0}{c_0^2}\dfrac{\partial^2\phi_1}{\partial r\partial t}\dfrac{\partial\phi_1}{\partial t}
- \dfrac{u_0}{c_0^2}\dfrac{\partial u_0}{\partial r}\dfrac{\partial\phi_1}{\partial r}\dfrac{\partial\phi_1}{\partial t}
- \dfrac{u_0^2}{c_0^2}\dfrac{\partial^2\phi_1}{\partial r^2}\dfrac{\partial\phi_1}{\partial t}
\nonumber
\\
- & \dfrac{u_0}{c_0^2}\dfrac{\partial\phi_1}{\partial r}\dfrac{\partial^2\phi_1}{\partial t^2}
- \dfrac{u_0}{c_0^2}\dfrac{\partial u_0}{\partial t}\left(\dfrac{\partial\phi_1}{\partial r}\right)^2
- \dfrac{u_0^2}{c_0^2}\dfrac{\partial^2\phi_1}{\partial r\partial t}\dfrac{\partial\phi_1}{\partial r}
\nonumber
\\
- & \dfrac{u_0^2}{c_0^2}\dfrac{\partial^2\phi_1}{\partial r\partial t}\dfrac{\partial\phi_1}{\partial r}
- \dfrac{u_0^2}{c_0^2}\dfrac{\partial u_0}{\partial r}\left(\dfrac{\partial\phi_1}{\partial r}\right)^2
- \dfrac{u_0^3}{c_0^2}\dfrac{\partial\phi_1}{\partial r}\dfrac{\partial^2\phi_1}{\partial r^2}.
\label{eq:DLDt2}
\end{align}
With Eq. \eqref{eq:Uterm}, the terms in Eq. \eqref{eq:DLDt2} can be rearranged as
\begin{align}
\dfrac{1}{\rho_0}\dfrac{\mathrm{D}\mathcal{L}}{\mathrm{D}t}
& = -\dfrac{1}{c_0^2}\left(\mathcal{U} + u_0^2\dfrac{\partial^2\phi_1}{\partial r^2}\right)\dfrac{\partial\phi_1}{\partial t}
- \dfrac{1}{c_0^2}\left(u_0\dfrac{\partial\phi_1}{\partial r} + \dfrac{\partial \phi_1}{\partial t}\right)\dfrac{\partial^2\phi_1}{\partial t^2}
\nonumber \\
& + u_0\left(1-\dfrac{u_0^2}{c_0^2}\right)\dfrac{\partial \phi_1}{\partial r}\dfrac{\partial^2\phi_1}{\partial r^2}
+ \left(\dfrac{\partial^2\phi_1}{\partial r\partial t} - \dfrac{u_0}{c_0^2}\mathcal{U}\right)\dfrac{\partial\phi_1}{\partial r}.
\label{eq:DLDt3}
\end{align}
Substituting Eqs. \eqref{eq:D2phi}, \eqref{eq:KD1phiD2phi}, \eqref{eq:gradphigradDphi}, and \eqref{eq:DLDt3} into Eq. \eqref{eq:convectiveKuznetsov2}, the convective lossless Kuznetsov equation is expressed in terms of partial derivatives, 
\begin{equation}
\mathcal{A}_1 \dfrac{\partial \phi_1}{\partial t}
+ \mathcal{A}_2 \dfrac{\partial^2 \phi_1}{\partial t^2}
+ \mathcal{K} \dfrac{\partial \phi_1}{\partial t}\dfrac{\partial^2 \phi_1}{\partial t^2}
+ \mathcal{U} 
+ \left(\mathcal{A}_G - \dfrac{c_0^2}{A}\dfrac{\partial A}{\partial r}\right) \dfrac{\partial \phi_1}{\partial r}
+ \mathcal{A}_L \dfrac{\partial^2 \phi_1}{\partial r^2} 
= 0,
\label{eq:convectiveKuznetsov_AAA}
\end{equation}
where the coefficients of Eq.~\eqref{eq:convectiveKuznetsov_AAA} are given by
\begin{align}
\mathcal{A}_1 &= \mathcal{K}\left(\mathcal{U} + u_0^2\, \dfrac{\partial^2\phi_1}{\partial r^2}\right),\\
\mathcal{A}_2 &= 1 + \mathcal{K} \, u_0\dfrac{\partial \phi_1}{\partial r},\\
\mathcal{A}_G &= \dfrac{\partial u_0}{\partial r}\dfrac{\partial \phi_1}{\partial r} + 2\dfrac{\partial^2 \phi_1}{\partial r\partial t} + \mathcal{K} \, u_0 \, \mathcal{U},\\
\mathcal{A}_L &= u_0^2 - c_0^2 + 2u_0\dfrac{\partial \phi_1}{\partial r} + \mathcal{K} \, u_0^3 \, \dfrac{\partial \phi_1}{\partial r}.
\end{align}
Following \citet{Solovchuk_et_al_2013}, the Newton linearization
\begin{equation}
\dfrac{\partial \phi_1}{\partial t}\dfrac{\partial^2 \phi_1}{\partial t^2} \approx 
\left(\dfrac{\partial \phi_1}{\partial t}\right)^o\dfrac{\partial^2 \phi_1}{\partial t^2} 
+
\left(\dfrac{\partial^2 \phi_1}{\partial t^2}\right)^o\dfrac{\partial \phi_1}{\partial t}
- \left(\dfrac{\partial \phi_1}{\partial t}\right)^o\left(\dfrac{\partial^2 \phi_1}{\partial t^2}\right)^o
\label{eq:NewtonLin}
\end{equation}
is applied to the nonlinear term in Eq.~\eqref{eq:convectiveKuznetsov_AAA}, where the superscript $o$ indicates the previous time level. With Eqs.~\eqref{eq:transDerivativeOp_space} and \eqref{eq:transDerivativeOp_time}, the partial derivatives are rewritten in the coordinates $\left(\xi,t\right)$ of the computational domain $\Theta$ as
\begin{align}
& \dfrac{\partial \phi_1}{\partial t} = \dfrac{\partial \Phi_1}{\partial t} + \dfrac{\partial \xi}{\partial t}\dfrac{\partial \Phi_1}{\partial \xi},
\label{eq:dPhidt}
\\
& \dfrac{\partial^2 \phi_1}{\partial t^2} = \dfrac{\partial^2 \Phi_1}{\partial t^2} + 2\dfrac{\partial \xi}{\partial t}\dfrac{\partial^2 \Phi_1}{\partial \xi\partial t} + \left(\dfrac{\partial^2 \xi}{\partial t^2} + \dfrac{\partial^2\xi}{\partial t \partial \xi}\right)\dfrac{\partial \Phi_1}{\partial \xi} + \left(\dfrac{\partial \xi}{\partial t}\right)^2\dfrac{\partial^2\Phi_1}{\partial\xi^2},
\label{eq:d2Phidt2}
\\
& \dfrac{\partial \phi_1}{\partial r} = \dfrac{\partial \xi}{\partial r}\dfrac{\partial \Phi_1}{\partial \xi},
\label{eq:dPhidr}
\\
& \dfrac{\partial^2 \phi_1}{\partial r^2} = \dfrac{\partial \xi}{\partial r}\dfrac{\partial^2\xi}{\partial x\partial\xi}\dfrac{\partial \Phi_1}{\partial \xi} + \left(\dfrac{\partial \xi}{\partial \xi}\right)^2\dfrac{\partial^2\Phi_1}{\partial\xi^2},
\label{eq:d2Phidr2}
\\
& \dfrac{\partial^2 \phi_1}{\partial r\partial t} = \dfrac{\partial \xi}{\partial r}\dfrac{\partial^2 \Phi_1}{\partial \xi\partial t} + \dfrac{\partial \xi}{\partial r}\dfrac{\partial^2\xi}{\partial t\partial \xi}\dfrac{\partial\Phi_1}{\partial\xi} + \dfrac{\partial \xi}{\partial t} \dfrac{\partial \xi}{\partial r}\dfrac{\partial^2\Phi_1}{\partial\xi^2}.
\label{eq:d2Phidrdt}
\end{align}
Applying Eqs. \eqref{eq:dPhidt}-\eqref{eq:d2Phidrdt} and Eq. \eqref{eq:NewtonLin}, Eq.~\eqref{eq:convectiveKuznetsov_AAA} becomes
\begin{align}
\mathcal{B}_1 \dfrac{\partial \Phi_1}{\partial t}
+ \mathcal{B}_2 \dfrac{\partial^2 \Phi_1}{\partial t^2}
+ 2\mathcal{B}_2 \dfrac{\partial \xi}{\partial t} \dfrac{\partial^2 \Phi_1}{\partial \xi\partial t}
- \mathcal{K}\left(\dfrac{\partial \phi_1}{\partial t}\right)^o\left(\dfrac{\partial^2 \phi_1}{\partial t^2}\right)^o
+ \mathcal{U} 
\nonumber \\
+ \left(\mathcal{B}_G - \dfrac{c_0^2}{A}\dfrac{\partial A}{\partial r} \dfrac{\partial \xi}{\partial r}\right)\dfrac{\partial \Phi_1}{\partial \xi} 
+ \mathcal{B}_L \dfrac{\partial^2 \Phi_1}{\partial \xi^2} 
= 0,
\label{eq:convectiveKuznetsov_BBB}
\end{align}
with the coefficients
\begin{align}
\mathcal{B}_1 &= \mathcal{A}_1 + \mathcal{K}\left(\dfrac{\partial^2 \phi_1}{\partial t^2}\right)^o,
\nonumber \\
\mathcal{B}_2 &= \mathcal{A}_2 + \mathcal{K}\left(\dfrac{\partial \phi_1}{\partial t}\right)^o,
\nonumber \\
\mathcal{B}_G &= \mathcal{B}_1\dfrac{\partial \xi}{\partial t} 
+ \mathcal{B}_2\left(\dfrac{\partial^2\xi}{\partial t^2} + \dfrac{\partial \xi}{\partial t} \dfrac{\partial^2\xi}{\partial t\partial \xi}\right) + \dfrac{\partial \xi}{\partial r}\left(\mathcal{A}_G
+ \mathcal{A}_L\dfrac{\partial^2\xi}{\partial r\partial \xi}\right),
\nonumber \\
\mathcal{B}_L &= \mathcal{B}_2\left(\dfrac{\partial \xi}{\partial t}\right)^2 + \mathcal{A}_L\left(\dfrac{\partial \xi}{\partial r}\right)^2.
\nonumber
\end{align}
The derivatives of $\phi_1$ in $\mathcal{A}_1$, $\mathcal{A}_2$, $\mathcal{A}_G$, $\mathcal{A}_L$ and $\mathcal{U}$ are given by Eqs.~\eqref{eq:transDerivativeOp_space} and \eqref{eq:transDerivativeOp_time}. The derivatives of $\xi$ follow from Eqs.~\eqref{eq:transDerivativeOp_space} and \eqref{eq:transDerivativeOp_time} and the linear transformation, Eq.~\eqref{eq:linTrans}, as \citep{Schenke_et_al_2022_PoF}
\begin{align}
\dfrac{\partial \xi}{\partial r} &= \dfrac{\mathcal{X}_{\mathrm{\infty}} - \mathcal{X}_{R}}{R_{\mathrm{stat}}-R\left(t\right)},
\label{eq:linJacobian} \\[4pt]
\dfrac{\partial \xi}{\partial t} &= 
-\dfrac{\mathcal{X}_{\mathrm{\infty}} - \xi}{R_{\mathrm{stat}} - R\left(t\right)} \dot R\left(t\right),
\label{eq:linq}
\\[4pt]
\dfrac{\partial^2 \xi}{\partial t \partial \xi} &= \dfrac{\dot R\left(t\right)}{R_{\mathrm{stat}} - R\left(t\right)},
\label{eq:lindivq} \\[4pt]
\dfrac{\partial^2 \xi}{\partial r \partial \xi} &= 0,
\label{eq:linJacobiandxi} \\[4pt]
\dfrac{\partial^2 \xi}{\partial t^2} &= 2\dfrac{\partial \xi}{\partial t}\dfrac{\partial^2\xi}{\partial t\partial \xi} - \ddot R\dfrac{\partial \xi}{\partial r} \dfrac{\mathcal{X}_{\infty} - \xi}{\mathcal{X}_{\infty} - \mathcal{X}_{R}}.
\label{eq:lindqdt}
\end{align}
Similar to our previous work \citep{Schenke_et_al_2022_PoF}, the functions describing the boundary motion associated with a stationary spherical background flow field with sonic transition at the sonic horizon $r_{\mathrm{h}}$ and initial position of the moving boundary $R_0$ are given by
\begin{align}
& R\left(t\right) = \left(R_0^{3} - 3c_0r_{\mathrm{h}}^2t\right)^{\frac{1}{3}},
\label{eq:R} \\[4pt]
& \dot R\left(t\right) = -c_0r_{\mathrm{h}}^2\left(R_0^{3} - 3c_0r_{\mathrm{h}}^2t\right)^{-\frac{2}{3}},
\label{eq:Rdot} \\[4pt]
& \ddot R\left(t\right) = -2c_0^2r_{\mathrm{h}}^{2n}\left(R_0^{3} - 3c_0r_{\mathrm{h}}^2t\right)^{-\frac{5}{3}}
\label{eq:Rddot}
\end{align}
for the acoustic black hole configuration, and by 
\begin{align}
& R\left(t\right) = \left(R_0^{3} + 3c_0r_{\mathrm{h}}^2t\right)^{\frac{1}{3}},
\label{eq:R_WH} \\[4pt]
& \dot R\left(t\right) = c_0r_{\mathrm{h}}^2\left(R_0^{3} + 3c_0r_{\mathrm{h}}^2t\right)^{-\frac{2}{3}},
\label{eq:Rdot_WH} \\[4pt]
& \ddot R\left(t\right) = -2c_0^2r_{\mathrm{h}}^{2n}\left(R_0^{3} + 3c_0r_{\mathrm{h}}^2t\right)^{-\frac{5}{3}}
\label{eq:Rddot_WH}
\end{align}
for the acoustic white hole configuration.

\section{Finite-difference discretization}\label{sec:fd}

Following the notation of \citet{Dey_and_Dey_1983} and \citet{Nascimento_et_al_2010}, with $j$ indicating the time instant and $i$ the grid point, the derivatives of $\Phi_1$ in Eq.~\eqref{eq:convectiveKuznetsov_BBB} are approximated by the finite differences
\begin{align}
\left.\dfrac{\partial^k \Phi_1}{\partial t^k}\right|^{j}_{i}
&=
\dfrac{a_0^{\left(k\right)}}{\Delta t^k} \Phi^{j+1}_{1,i} 
+ 
\dfrac{1}{\Delta t^k}\underbrace{\sum_{n=1}^{N-1} a_n^{\left(k\right)}\Phi^{j+1-n}_{1,i}
}_{b^{\left(k,j\right)}_i},
\label{eq:fddt_compact} \\
\left.\dfrac{\partial^k \Phi_1}{\partial \xi^k}\right|^{j}_{i}
&=
\dfrac{\displaystyle a_0^{\left(k\right)}\Phi^{j}_{1,i} + \sum_{n=1}^{\left(N-1\right)/2} \left( a_n^{\left(k\right)}\Phi^{j}_{1,i+n} + a_{n}^{\left(k\right)}\Phi^{j}_{1,i-n}\right)}{\Delta\xi^k},
\label{eq:fddxi_compact}
\end{align}
where $a_n$ denotes the corresponding finite-difference coefficients. Analogous to the spatial derivatives, the mixed spatial-temporal derivatives are evaluated at the time-level $j$. This treatment of the mixed derivative term $\left(\partial \xi/\partial t\right)\partial^2\Phi_1/\partial \xi\partial t$ in Eq.~\eqref{eq:convectiveKuznetsov_BBB} has a diffusive effect on the numerical solution in cases where the physical domain $\Omega\left(t\right)$ is subject to significant deformations. The numerical diffusion is counteracted by the blending of two analytically-equivalent expressions, where
\begin{equation}
\dfrac{\partial \xi}{\partial t} \dfrac{\partial^2 \Phi_1}{\partial \xi\partial t}
= \dfrac{1}{2}\dfrac{\partial \xi}{\partial t} \dfrac{\partial^2 \Phi_1}{\partial \xi\partial t} + 
\dfrac{1}{2}\left[
\dfrac{\partial^2}{\partial \xi\partial t}\left(\dfrac{\partial \xi}{\partial t}\Phi_1\right)
- \dfrac{\partial^2 \xi}{\partial t^2 \partial \xi} \dfrac{\partial \Phi_1}{\partial \xi}
- \dfrac{\partial^3 \xi}{\partial t^2\partial \xi}\Phi_1
- \dfrac{\partial^2 \xi}{\partial t \partial \xi} \dfrac{\partial \Phi_1}{\partial t}\right],
\label{eq:mixedTerm}
\end{equation}
which is discretized with the above schemes. With $b^{\left(k,j\right)}_i$ specified in Eq.~\eqref{eq:fddt_compact}, the discretized wave equation is
\begin{equation}
\widetilde{\Phi}_{1,i}\overbrace{\sum_{k=1}^2\mathcal{A}_{k,i}^j \dfrac{a_0^{\left(k\right)}}{\Delta t^k}}^{\mathcal{H}^j_i}
+
\overbrace{\mathcal{R}^j_i
+
\sum_{k=1}^2\mathcal{A}_{k,i}^j \dfrac{b^{\left(k,j\right)}_i}{\Delta t^k}}^{\mathcal{S}^j_i}
= 
- \mathcal{A}^j_{L,i}\left(\dfrac{\partial^2 \Phi_1}{\partial \xi^2}\right)^j_i
+ \left(\dfrac{c_0^2}{A}\dfrac{\partial A}{\partial r} \dfrac{\partial \xi}{\partial r}\right)^j_i\left(\dfrac{\partial \Phi_1}{\partial \xi}\right)^j_i,
\label{eq:predictor} 
\end{equation}
where $\mathcal{R}$ includes all mixed spatial-temporal derivatives as well as the gradient terms that are not associated with the Laplacian on the right-hand side of Eq.~\eqref{eq:predictor}.
Eq.~\eqref{eq:predictor} is solved explicitly for $\widetilde{\Phi}_{1,i}$. In order to stabilize the solution against dispersive numerical noise, $\widetilde{\Phi}_{1,i}$ is treated as an interim (predicted) solution, which is subsequently corrected to obtain the new solution $\Phi^{j+1}_{1,i}$ at time level $j+1$. With $1-\gamma$ and $\gamma$ being the weights of the predicted and corrected solutions, respectively, the corrector step is given as \citep{Dey_and_Dey_1983, Nascimento_et_al_2010}
\begin{equation}
\Phi^{j+1}_{1,i} = \left(1-\gamma\right)\widetilde{\Phi}_{1,i} + \dfrac{\gamma}{\mathcal{H}^j_i}\left[\mathcal{A}^j_{L,i}\left(\dfrac{\partial^2\widetilde{\Phi}_1}{\partial \xi^2}\right)_i - \left(\dfrac{c_0^2}{A}\dfrac{\partial A}{\partial r} \dfrac{\partial \xi}{\partial r}\right)^j_i\left(\dfrac{\partial \widetilde{\Phi}_1}{\partial \xi}\right)_i + \mathcal{S}^j_i
\right].
\label{eq:corrector}
\end{equation}
Prior to the corrector step, the Laplacian on the right-hand side of Eq.~\eqref{eq:predictor}, also involving the gradient contribution related to the flow geometry, is reevaluated based on the interim solution $\widetilde{\Phi}_1$. At $r=R_{\mathrm{stat}}$, an absorbing boundary condition is imposed to allow the acoustic waves to pass the domain boundary. We use Mur's first order absorbing boundary condition \citep{Mur_1981}, which requires the perturbation potential to satisfy
\begin{equation}
\Phi_{1,N}^{j+1} = \Phi_{1,N-1}^{j} + \dfrac{\mathrm{CFL-1}}{\mathrm{CFL}+1}\left(\Phi_{1,N-1}^{j+1}-\Phi_{1,N}^{j}\right),
\label{eq:Mur}
\end{equation}
where the Courant-Friedrichs-Lewy number $\mathrm{CFL}$, expressed in terms of the speed of sound $c_0$ and the step size $\Delta x$ in the physical domain $\Omega\left(t\right)$, is given by
\begin{equation}
\mathrm{CFL} = \dfrac{c_0\Delta t}{\Delta x}.
\label{eq:CFL}
\end{equation}
The finite-difference coefficients for the temporal and the spatial derivatives are given in Tables \ref{tab:fddtCoeff} and \ref{tab:fddxiCoeff}, respectively. We opt for a first-order approximation of the temporal derivatives for two reasons. First, the scheme corresponds to the original predictor-corrector method by \citet{Dey_and_Dey_1983} and, second, the scheme is compatible with Mur's first-order absorbing boundary condition \citep{Mur_1981}.

\begin{table}
\vspace{1em}
\centering
\begin{minipage}{0.4\textwidth}
\centering
\caption{Coefficients of the finite-difference approximations of the $k$-th temporal derivative.}
\label{tab:fddtCoeff}
\begin{tabular}{ r | c c c }
$k$ & $a_{0}$ & $a_{1}$ & $a_{2}$ \\[2pt]
\hline
$1$ & $1$ & $-1$ & $0$ \\[2pt]
$2$ & $1$ & $-2$ & $1$ \\[2pt]
\end{tabular}
\end{minipage}
\qquad
\begin{minipage}{0.4\textwidth}
\centering
\caption{Coefficients of the finite-difference approximations of the $k$-th spatial derivative.}
\label{tab:fddxiCoeff}
\begin{tabular}{ r | c c c }
$k$ & $a_{-1}$ & $a_{0}$ & $a_{1}$ \\[2pt]
\hline
$1$ & $-1/2$ & $0$ & $1/2$ \\[2pt]
$2$ & $1$ & $-2$ & $1$  \\[2pt]
\end{tabular}
\end{minipage}
\end{table}

\section{Transformation between the acoustic pressure and the perturbation potential}\label{sec:Transformation between the acoustic pressure and the perturbation potential}

While the solution for $\Phi_1\left(\xi,t\right)=\phi_1\left(r\left(\xi,t\right),t\right)$ is the primary result of the numerical simulation, we are ultimately interested in the solution for the acoustic pressure $p_1$, which requires to specify the transformation between these two quantities within the numerical framework presented in Sec.~\ref{sec:fd}. This transformation is further needed in order to specify the boundary condition for $\Phi_1\left(\xi=0,t\right)=\phi_1\left(r=R\left(t\right),t\right)$ at the moving wave-emitting boundary based on the boundary value for the acoustic pressure, i.e. the acoustic excitation pressure $p_1\left(r=R\left(t\right),t\right)$. The transformation between $\Phi_1$ and $p_1$ is achieved by transforming Eq.~\eqref{eq:rhoExpansion} from the moving physical domain $\Omega\left(t\right)$ into the fixed computational domain $\Theta$, using the techniques presented in Sec.~\ref{sec:coordinatetransformation}, and by subsequently discretizing the equation using the finite difference approximations presented in Sec.~\ref{sec:fd}.

Substituting the potential form of the Lagrangian density $\mathcal{L}$, given by Eq.~\eqref{eq:LagrangianDensity2}, into Eq.~\eqref{eq:rhoExpansion} gives
\begin{equation}
\dfrac{p_1}{\rho_0} = c_0^2\dfrac{\rho_1}{\rho_0} = - \dfrac{\mathrm{D}\phi_1}{\mathrm{D}t} + \dfrac{1}{2}\left(\dfrac{\partial\phi_1}{\partial r}\right)^2 + \dfrac{3 - 2\beta}{2c_0^2}\left(\dfrac{\mathrm{D}\phi_1}{\mathrm{D}t}\right)^2.
\label{eq:rhoExpansion_b}
\end{equation}
With the definition of the material derivative operator given by Eq.~\eqref{eq:MaterialDerivative} and the Newton linearization \citep{Solovchuk_et_al_2013}
\begin{equation}
\left(\dfrac{\partial \phi_1}{\partial t}\right)^2 \approx 2\left(\dfrac{\partial \phi_1}{\partial t}\right)^o\dfrac{\partial \phi_1}{\partial t} - \left[\left(\dfrac{\partial \phi_1}{\partial t}\right)^o\right]^2,
\label{eq:NewtonLin2}
\end{equation}
Eq.~\eqref{eq:rhoExpansion} can be rewritten as
\begin{equation}
\underbrace{\left[1-\mathcal{K}\left(\left(\dfrac{\partial \phi_1}{\partial t}\right)^o - u_0\dfrac{\partial \phi_1}{\partial r}\right)\right]}_{\mathcal{A}_1}\dfrac{\partial \phi_1}{\partial t} 
+ 
\underbrace{\left[u_0 + \dfrac{1}{2}\left(1-\mathcal{K}u_0^2\right)\dfrac{\partial \phi_1}{\partial r}\right]}_{\mathcal{A}_G}\dfrac{\partial \phi_1}{\partial r}
+
\underbrace{\dfrac{\mathcal{K}}{2}\left[\left(\dfrac{\partial \phi_1}{\partial t}\right)^o\right]^2}_{\mathcal{N}}
+
\dfrac{p_1}{\rho_0} 
= 
0,
\label{eq:rhoExpansion2}
\end{equation}
where $\mathcal{K} = \left(3-2\beta\right)/c_0^2$.
Applying the change of variables as specified by Eqs.~\eqref{eq:transDerivativeOp_space} and \eqref{eq:transDerivativeOp_time} to Eq.~\eqref{eq:rhoExpansion2}, we get
\begin{equation}
\mathcal{A}_1\dfrac{\partial \Phi_1}{\partial t}
+
\left(\mathcal{A}_1\dfrac{\partial \xi}{\partial t} + \mathcal{A}_G \dfrac{\partial \xi}{\partial r}\right)\dfrac{\partial \Phi_1}{\partial \xi}
+
\mathcal{N} + \dfrac{p_1}{\rho_0} = 0,
\label{eq:rhoExpansion_transformed}
\end{equation}
with the coefficients
\begin{align*}
\mathcal{A}_1 &= 1-\mathcal{K}\left(\left(\dfrac{\partial \phi_1}{\partial t}\right)^o - u_0\dfrac{\partial \phi_1}{\partial r}\right),
\\
\mathcal{A}_G &= u_0 + \dfrac{1}{2}\left(1-\mathcal{K}u_0^2\right)\dfrac{\partial \phi_1}{\partial r},
\\
\mathcal{N} &= \dfrac{\mathcal{K}}{2}\left[\left(\dfrac{\partial \phi_1}{\partial t}\right)^o\right]^2
\\
\mathcal{K} &= \dfrac{3 - 2\beta}{c_0^2}.
\end{align*}
The derivatives of $\phi_1$ in $\mathcal{A}_1$, $\mathcal{A}_G$, and $\mathcal{N}$ are again given by Eqs.~\eqref{eq:transDerivativeOp_space} and \eqref{eq:transDerivativeOp_time}, analogously to the step following Eq.~\eqref{eq:convectiveKuznetsov_BBB}. Since the coefficients only involve either spatial derivatives or the old solution of the partial time derivative as indicated by the superscript $o$, they do not contain values for $\phi_1\left(r\left(\xi,t\right),t\right)=\Phi_1\left(\xi,t\right)$ at the new time level $j+1$ and can, therefore, be reconstructed from the known solution available at time level $j$. With the finite difference approximation of the partial time derivative given by Eq.~\eqref{eq:fddt_compact}, the semi-discrete form of Eq.~\eqref{eq:rhoExpansion_transformed} becomes
\begin{equation}
\mathcal{A}_1\dfrac{a_0^{\left(1\right)}\Phi_{1,i}^{j+1} + b^{\left(1,j\right)}_i}{\Delta t}
=
-\left[
\left(\mathcal{A}_1\dfrac{\partial \xi}{\partial t} + \mathcal{A}_G \dfrac{\partial \xi}{\partial r}\right)\dfrac{\partial \Phi_1}{\partial \xi}
+
\mathcal{N} + \dfrac{p_1}{\rho_0}
\right]_i^j.
\label{eq:rhoExpansion_discr}
\end{equation}
Eq.~\eqref{eq:rhoExpansion_discr} is used to (i) reconstruct the acoustic pressure field $p_{1,i}^j$ once $\Phi_{1,i}^{j+1}$ has been computed, and (ii) to convert the acoustic excitation pressure $p_{1,i=0}^j$ at the moving wave-emitting domain boundary into the corresponding excitation potential $\Phi_{1,i=0}^{j}$.

\fancyhf{}
\fancyhead[r]{Schenke et al.: Amplitude modulation of acoustic waves ... \hfill Supplementary material}
\fancyfoot[c]{\thepage}
\newcommand{\noop}[1]{}